\documentclass[12pt]{iopart}
\usepackage{iopams}
\usepackage{graphicx}
\usepackage{amssymb}

\begin{document}

\title[First hitting times of non-backtracking RWs on ER networks]
{The distribution of first hitting times \\ 
of non-backtracking random walks \\
on Erd\H{o}s-R\'enyi networks
}

\author{Ido Tishby, Ofer Biham and Eytan Katzav}
\address{Racah Institute of Physics, 
The Hebrew University, Jerusalem 91904, Israel.}
\eads{\mailto{ido.tishby@mail.huji.ac.il}, \mailto{biham@phys.huji.ac.il}, 
\mailto{eytan.katzav@mail.huji.ac.il}}

\begin{abstract}
We present analytical results for 
the distribution of first hitting times of
non-backtracking random walks on finite
Erd\H{o}s-R\'enyi networks of $N$ nodes.
The walkers hop randomly between adjacent nodes 
on the network,
without stepping back to the previous node,
until they hit a node which they have already visited before
or get trapped in a dead-end node.
At this point, the path is terminated.
The length, $d$,
of the resulting path,
is called the first hitting time.
Using recursion equations, we obtain 
analytical results for the tail distribution of first hitting times, 
$P(d > \ell)$, 
$\ell=0,1,2,\dots$,
of non-backtracking random walks starting from a
random initial node.
It turns out that the distribution
$P(d > \ell)$ 
is given by a product of a
discrete Rayleigh distribution and an exponential distribution.
We obtain analytical expressions for 
central measures (mean and median) and
a dispersion measure (standard deviation) 
of this distribution.
It is found that the paths of non-backtracking 
random walks, up to their termination at the first hitting time,
are longer, on average, than those of the corresponding simple
random walks.
However, they are shorter than those of  
self avoiding walks 
on the same network,
which terminate at the last hitting time.
We obtain analytical results for the probabilities,
$p_{\rm ret}$ and $p_{\rm trap}$, 
that a path will terminate 
by retracing, namely stepping into an already visited node,
or by trapping, namely entering a node 
of degree $k=1$, which has no exit link,
respectively.
It is shown that in dilute networks the dominant termination
scenario is trapping while in dense networks
most paths terminate by retracing.
We obtain expressions for the 
conditional tail distributions of path lengths,
$P(d>\ell | {\rm ret})$ 
and
$P(d>\ell | \rm{trap})$,
for those paths which terminate by
retracing or by trapping, respectively.
We also study a class of generalized non-backtracking random 
walk models which not only avoid the backtracking step into the previous node
but avoid stepping into the last $S$ visited nodes, where
$S=2,3,\dots,N-2$.
Note that the case of $S=1$ coincides with
the non-backtracking random walk model described above,
while the case of $S=N-1$ coincides
with the self avoiding walk.  
\end{abstract}


\vspace{2pc}

\submitto{\JPA (\today)}
\date
\maketitle

\section{Introduction}

Random walk (RW) models 
\cite{Spitzer1964,Weiss1994}
are useful for the study of a large variety of
stochastic processes 
such as diffusion
\cite{Berg1993,Ibe2013},
polymer structure 
\cite{Fisher1966,Degennes1979}
and random search
\cite{Evans2011,Lopez2012}.
These models were studied extensively 
in geometries including
continuous space
\cite{Lawler2010b}, 
regular lattices 
\cite{Lawler2010a},
fractals 
\cite{ben-Avraham2000}
and 
random networks
\cite{Noh2004}.
A RW on a network hops randomly at each time step
to one of the nodes which are adjacent to the current node.
Thus, if the current node is of degree $k$, 
the probability of each one of its neighbors to be 
selected by the RW 
is $1/k$.
In some of the steps the RW hops into new nodes, 
which have not been visited before.
In other steps it 
may backtrack its path into the previous node or
hop into nodes already visited at earlier times.
It was found that RWs on random networks
are highly effective in exploring the network, retracing their
steps much less frequently than RWs on low dimensional lattices
\cite{Montroll1965}.
Recent studies 
of random walks on random networks 
produced analytical results for the
mean first passage time 
between random pairs of nodes
\cite{Sood2005}, 
the average number of distinct nodes 
visited by a RW after $t$ time steps
\cite{Debacco2015}
and the mean cover time
\cite{Kahn1989}.

A special type of random walk model, 
which has been studied 
extensively on regular lattices, 
is the self avoiding walk
(SAW).
This is a random walk which does not 
visit the same node more than once
\cite{Madras1996}.
At each time step, the walker chooses its next move
randomly from the neighbors 
of its present node, 
excluding nodes
that were already visited.
The path terminates when the RW
reaches a dead-end node from which it cannot exit, 
namely a node which does not 
have any yet unvisited neighbors.
The length of the path, 
$d$,
is given by the number 
of steps made until the path has terminated.
The path length of an SAW on a connected network of size $N$
can take values between $1$ and $N-1$.
The latter case corresponds to 
a Hamiltonian path
\cite{Bollobas2001}.
More specifically, the SAW path lengths between 
a given pair of nodes,
$i$ and $j$, are distributed in the range bounded from
below by the shortest path length between these nodes
\cite{Katzav2015,Nitzan2016}
and 
from above by
the longest non-overlapping path between them
\cite{Karger1997}.
The path length of an SAW on a random network
is called the
{\it last hitting time}
\cite{Herrero2005}.
In Ref.
\cite{Tishby2016} 
we presented analytical results for the distribution of SAW
path lengths, or
last hitting times, 
on 
ER
networks
\cite{Erdos1959,Erdos1960,Erdos1961}.
These 
SAW paths are often referred to as kinetic 
growth self-avoiding walks 
\cite{Herrero2007}, 
or true self avoiding
walks 
\cite{Slade2011}.
This is 
in contrast to SAW paths which are uniformly sampled among
all possible self avoiding paths of a given length.
It was found that
the distribution of SAW path lengths follows a discrete version of
the Gompertz distribution
\cite{Gompertz1825,Johnson1995,Shklovskii2005,Ohishi2008}.
This means that the SAWs exhibit a termination rate per step which
increases exponentially with the number of steps already pursued.

Another important time scale which appears 
in random walks on networks is the
{\it first hitting time}
\cite{Debacco2015},
also referred to as the first intersection length
\cite{Herrero2003,Herrero2005b}.
This time scale emerges in a class of RW models which are not 
restricted to be self avoiding. In these models the RW hops freely
between adjacent nodes until it enters a node which has already been visited
before. At this point the path is terminated. 
The number of time steps up to termination of the path, 
which coincides with the path length, 
is called the first hitting time.
This time, on average, is much smaller than the last hitting 
time, namely the path length of an 
SAW on the same network.
This is due to the fact that the RW path may terminate at any 
time step, $t>1$, by randomly hopping into an
already visited node, even if the current node has one or more 
yet-unvisited neighbors.
This is in contrast with the SAW path, which
terminates only when the current node does not have any 
yet-unvisited neighbors.
In Ref.
\cite{Tishby2016b}
we presented
analytical results for the
distribution 
of first hitting times of RWs on ER networks.
In the analysis, we utilized the fact that up to its 
termination the RW follows an
SAW path. 
The path pursued by the RW may terminate either by 
backtracking into the previous node
or by retracing itself, namely stepping into a node which was 
already visited two or more time steps earlier.
By calculating the probabilities of these two termination scenarios, we
obtained analytical results for the distribution of 
first hitting times of RWs on ER networks.
We also obtained analytical results for the probabilities,
$p_{\rm backtrack}$ and $p_{\rm ret}$,
that a RW, starting from a random initial node,
will terminate by backtracking or by retracing,
respectively.
It was found that in dilute networks most paths terminate
by backtracking while in dense networks most paths 
terminate by retracing.

The RW model 
studied in Ref.
\cite{Tishby2016b}
and the SAW model studied in Ref.
\cite{Tishby2016}
are very different from each other. 
The SAW may be considered as a walker which maintains a complete
record of all the nodes it has visited and avoids 
stepping into any of them again.
On the other hand, the RW does not keep track of its path and thus may 
hop into an already visited node at any time step.
This difference leads to different
termination scenarios in the two models. 
While the RW path is terminated by either backtracking or by retracing of
its own path, the SAW terminates only when it reaches a dead end
node which does have any yet unvisited neighbor.
The RW model and the SAW model can be considered as two 
opposite limits in a 
class of RW models 
which keep track of the last $S$ visited nodes, where
$S=0, 1, \dots, N-1$,
and avoid stepping into any of them again.
Put differently,
in the $t+1$ time step such RW avoids hopping 
from the current node, to which entered at time $t$,
into any of the nodes visited at
times $t-1$, $t-2$, $\dots$, $t-S$,
even if they are adjacent to the current node.
However, it may hop into nodes
visited at earlier times, in which case the path is terminated.
The case of $S=0$ corresponds to the RW studied in Ref.
\cite{Tishby2016b}.
The case of $S=1$ corresponds to the non-backtracking random walk (NBW),
which avoids hopping back into the previous node.
RW models with $S>1$ are called generalized NBW models, 
which are closely related to a class of models known as tourist walks
\cite{Stanley2001,Silva2016}.
Note that the case of $S=N-1$ coincides with the SAW model studied in Ref.
\cite{Tishby2016}.

The paths of
NBWs have been studied on regular lattices and random graphs
\cite{Alon2007}.
It was shown that they explore the network more efficiently than
RWs. It was also shown that they mix faster,
namely require a shorter transient time
to reach the stationary distribution of visiting frequencies
throughout the network.
The path of the NBW model studied here
may terminate either by retracing, namely by hopping into  
a node which has already been visited before,
or by trapping, namely entering a dead-end node from which it cannot exit.

NBWs provide a useful description of a large variety of 
randomly wandering objects on networks. These objects 
are endowed with a memory, which enables them to avoid 
revisiting the nodes visited in the last $S$ time steps.
The non-backtracking property is often highly beneficial.
For example, it speeds-up the performance of web-crawlers
which constantly scan the world-wide-web. Also, it extends
the life expectancy of random foragers which feed on the
network.

Web crawlers are robots which scan the internet and 
assemble information from webpages, to be used in 
search engines and other databases 
\cite{Kobayashi2000}.
They hop randomly
in the web following the hyperlinks. To optimize the
efficiency of web crawlers it is important to avoid
revisiting sites at too high frequency. On the other 
hand, web crawlers need to keep the information fresh,
namely revisit web sites frequently enough to account
for updates. A generalized NBW protocol provides a
strategy for such web crawlers, because  
tuning the parameter $S$ enables to balance between
these competing requirements.

To make the connection to foraging theory,
consider an animal, which is 
randomly foraging in a random
network environment. 
Each time the animal visits 
a node it consumes all the food
available in this node and needs to move on to 
one of the adjacent nodes.
The model describes rather harsh conditions, in 
which the regeneration of 
resources is very slow and the visited nodes do 
not replenish within the lifetime
of the forager. Moreover, the forager does not 
carry any reserves and in order
to survive it must hit a vital node each and every time. 
Under these conditions, the distribution of
life expectancies of the foragers coincides with the
distribution of first hitting times.
Clearly, a forager which avoids hopping back to the
previous node (thus described by an NBW model)
will have a higher life expectancy than a forager
which may hop back to the previous node
(described by an RW model).
Foragers which avoid revisiting the last $S$ visited
nodes are described by the generalized NBW model.
Their life expectancy further increases as $S$ is
increased.

Several variants of the forager model have been 
studied on lattices of 
different dimensions. 
For example, the case in which the forager
carries sufficient resources which enable it to 
avoid starvation even when it 
visits several non-replenished 
nodes in a row, was recently studied 
\cite{Benichou2014,Chupeau2016}.
The effect of the regeneration rate of the visited
nodes was also examined.
It was shown that under slow regeneration
the forager is susceptible to starvation, 
while above some threshold regeneration 
rate, the probability of starvation diminishes
\cite{Chupeau2016a}. 

In this paper we present analytical results 
for the distribution of first hitting times
of NBWs on ER networks.
We obtain expressions for the mean, 
median and standard deviation
of this distribution in terms of the parameters of the network.
It turns out that the termination of an NBW path may occur
either by the retracing scenario or by trapping 
in a dead-end (leaf) node, from which it cannot exit.
We obtain analytical results for the probabilities,
$p_{\rm ret}$ and $p_{\rm trap}$,
that an NBW
will terminate by retracing or by trapping,
respectively.
It is found that in dilute networks most paths terminate
by trapping while in dense networks most paths 
terminate by retracing.
We also obtain expressions 
for the
conditional tail distributions of path lengths,
$P(d>\ell | {\rm ret})$ 
and
$P(d>\ell | {\rm trap})$,
given that the NBWs are terminated by
retracing or by trapping,
respectively.
We show that as $S$ is increased, the termination 
probability decreases and
the mean path length increases.

The paper is organized as follows.
In Sec. 2 we describe the class of non-backtracking random walk models
on the ER network. 
In Sec. 3 we consider the 
evolution of the subnetwork which consists of the yet unvisited nodes.
In Sec. 4 we derive analytical results for the distribution 
of path lengths, or first hitting times, of NBWs on ER networks.
In Sec. 5 we obtain analytical expressions for two central measures
(mean and median) and a dispersion measure (standard deviation) 
of this distribution.
In Sec. 6 we analyze the distributions of path lengths of NBWs,
conditioned on the termination scenario.
In Sec. 7 we consider generalized NBW models 
and present analytical results for the distribution 
of first hitting times in these models.
The results are summarized and discussed in Sec. 8.
The details of the calculations of $P(d>\ell)$ for the NBW model
and for the generalized NBW model are presented in Appendices A and B,
respectively. The calculation of the mean of $P(d>\ell)$ for the generalized
NBW model is presented in Appendix C.

\section{The non-backtracking random walk models}

Consider a random walk on a random network of $N$ nodes,
starting from a random initial node.
Each time step the walker chooses randomly one of the neighbors
of its current node, and hops to the chosen node. 
The RW continues to hop between adjacent nodes
as long as it does not visit any node more than once.
For concreteness, we denote the initial node by $x_0$ and
the subsequent nodes along the path by
$x_t$, $t=1,2,\dots$, where $x_t$ is the node which
the RW enters at time $t$ and leaves at time $t+1$.
The RW path is
terminated upon the first time it steps into an already visited node.
The resulting path length
is referred to as the first hitting time
or the first intersection length.

The NBW may hop randomly to
any neighbor of the current node, 
except for the previous node, from which it entered the current node.
This significantly reduces the termination probability because the 
previous node is the only node
which is guaranteed to be connected to the current node.
The NBW exhibits two termination scenarios, namely
retracing and trapping.
In the retracing scenario, the NBW hops into an already visited node. 
In the trapping scenario it hops into a dead end (leaf) node
of degree $k=1$, from which the only way out is by backtracking,
which is not allowed.

We also consider generalized NBW models, 
which keep track of the last $S$ visited
nodes and avoid hopping into them.
The parameter, $S$, may take values in the range
$1 \le S \le N-1$.
The case of $S=1$ is the NBW model,
in which backtracking into the previous node is not allowed.
The paths of NBWs may terminate
either by hopping into nodes already visited at earlier times (retracing)
or by trapping in a leaf node of degree $k=1$ from which they
cannot exit.
Generalized NBW models, with
$S>1$, avoid hopping into
any of the nodes visited at times $t=1$, $t-2$, $\dots$, $t-S$.
The paths of generalized NBWs may terminate either by retracing 
or by trapping in a node which
is surrounded by the tail of $S$ nodes which they cannot enter. 
The case of $S=N-1$ coincides with the SAW model.

\section{Evolution of the subnetwork of the yet-unvisited nodes}

Consider an $ER(N,p)$ network.
The degree $k_i$ of node $i=1,\dots,N$ 
is the number of links connected to this node.
The degree distribution 
$p(k)$
of the ER network 
is a binomial distribution, 
which in the sparse limit
($p \ll 1$)
is approximated by a Poisson distribution of the form 

\begin{equation}
p(k)=\frac{{c}^{k}}{k!}e^{-c},
\label{eq:poisson}
\end{equation}

\noindent
where
$c=(N-1)p$
is the average degree.
In the asymptotic limit 
($N \rightarrow \infty$),
the ER
network exhibits a phase transition at 
$c=1$ (a percolation transition), such that for
$c<1$
the network consists only of small clusters and isolated nodes, while for 
$c>1$
there is a giant cluster which includes a macroscopic fraction of the network, in addition
to small clusters and isolated nodes. 
At 
$c = \ln N$, 
there is a second transition, 
above which the entire network is included in
the giant cluster and there are no isolated components. 

Here we focus on the regime
above the percolation transition, namely 
$c>1$.
For
$1 < c < \ln N$
the fraction of
isolated nodes among all nodes in the network is given by
$i(c) = \exp(-c)$.
In order to avoid the trivial case of an NBW starting
on an isolated node, we performed the analysis presented
below for the case in which the initial node is non-isolated.
At time steps $t=2,\dots,S+1$ the generalized NBW avoids hopping into any
of the previously visited nodes, and thus behaves as an SAW. 
At $t \ge S+2$, it avoids hopping
from the node it entered at time $t-1$ into the nodes visited at times
$t-2$, $t-3$, $\dots$, $t-S-1$ but may hop into nodes visited at
earlier times, thus causing termination of the path.
In these models, as long as the path does not terminate, it is
identical to an SAW path.

The NBW path divides the network into two subnetworks,
one consists of the already visited nodes and the
other consists of the yet-unvisited nodes. 
After $t$ time steps the size of the subnetwork of 
visited nodes is $t+1$ 
(including the initial node),
while the size of the network of
yet-unvisited nodes is
$N(t)=N-t-1$. 
We denote the degree distribution of the subnetwork 
of the yet-unvisited nodes at time $t$ by
$p_t(k)$, $k=0,\dots,N(t)-1$,
where
$p_0(k)=p(k)$,
namely the original degree distribution.
The average degree of this subnetwork 
is given by

\begin{equation}
\langle k \rangle_t = \sum_{k=0}^{N(t)-1} k p_t(k).
\label{eq:<k>}
\end{equation}

\noindent
We denote it by
$c(t) = \langle k \rangle_t$,
where $c(0)=c$.

RWs on random networks exhibit higher probabilities of visiting nodes of high degrees. 
More precisely, 
the probability that in a given time step a RW will visit a node 
of degree $k$,
is given by  
$k p(k)/c$,
namely it is proportional to the 
degree of the node.
A special property of the 
Poisson distribution is that
the probability
${k p(k)}/{c} = p(k-1)$.
This means that,
the probability that the node
visited at time $t+1$ will be of  
degree $k$ 
is given by 
$p(k-1)$. 
In Ref.
\cite{Tishby2016}
it was shown that the degree distribution of the
subnetwork which consists of the yet unvisited nodes 
at time $t$ is 

\begin{equation}
p_t(k) = \frac{c(t)^k}{k!}e^{-c(t)},
\label{eq:p_t(k)}
\end{equation}

\noindent
where 

\begin{equation}
c(t) = \left(1-\frac{t}{N-1}\right)c
\label{eq:coft}
\end{equation}

\noindent
is the mean degree of this subnetwork.
This means that the subnetwork of the yet-unvisited nodes
remains an ER network, while its size and mean degree decrease linearly
in time. 
Therefore, the degree distribution of this 
subnetwork also satisfies
${k p_{t}(k)}/{c(t)} = p_{t}(k-1)$.

\section{The distribution of first hitting times of the NBW model}

Consider an NBW
(with $S=1$) 
on an ER network of $N$ nodes.
The NBW starts from a random node with degree
$k \ge 1$ (non-isolated node) and hops 
randomly between nearest neighbor nodes
without backtracking into the previous node.
The path of the NBW may terminate either by the
retracing scenario or by the trapping scenario.
In the retracing mechanism the NBW steps into
a node which was already visited two or more time steps earlier.
In the trapping mechanism the NBW enters a dead end (leaf) 
node of degree $k=1$ from which it cannot exit.

In case that the NBW has pursued $t$ steps, without retracing
its path and without getting trapped, 
the path length, $d$, is guaranteed to satisfy $d > t-1$.
At this point, the probability that the path will not
terminate in the $t+1$ step is denoted 
by the conditional probability
$P(d>t|d>t-1)$. 
This conditional probability can be expressed as a product
of the form

\begin{equation} 
P(d > t|d > t-1) =  P_{\rm ret}(d > t|d > t-1) P_{\rm trap}(d > t|d > t-1).  
\label{eq:cond3}
\end{equation}

\noindent
The conditional probability
$P_{\rm trap}(d > t|d > t-1)$ 
is the probability that the NBW will 
not terminate by the trapping scenario at the $t+1$ time step.
Given that the NBW is not terminated by trapping at the $t+1$ step,
the conditional probability
$P_{\rm ret}(d > t|d > t-1)$ 
is the probability that it will also
not terminate by the retracing scenario, namely that it will not 
step into a node already visited at an earlier time.

The probability that the NBW will not terminate by the trapping mechanism
in the $t+1$ time step is given by the probability that the
node it entered 
at time $t$
is of degree $k>1$.
This probability is given by

\begin{equation}
P_{\rm trap}(d > t|d > t-1) =
\sum_{k=2}^{N-1} \frac{k p(k)}{c}.
\label{eq:trap2}
\end{equation}

\noindent
Inserting in 
Eq. 
(\ref{eq:trap2})
the Poisson distribution 
of Eq.
(\ref{eq:poisson})
we obtain 

\begin{equation}
P_{\rm trap}(d > t|d > t-1) =
1 - e^{-c}.
\label{eq:trap3}
\end{equation}

\noindent
Note that this probability does not depend on the time, $t$.
Given that the NBW was not terminated by trapping at the
$t+1$ time step, we will now evaluate the probability,
$P_{\rm ret}(d > t|d > t-1)$, 
that it will also not terminate by retracing.
Apart from the current node and the previous node, there are $N-2$
possible nodes which may be connected to the current node, 
each one of them with probability $p$.
The fact that the possibility of trapping
was already eliminated for the $t+1$ step,
guarantees that at least one of these $N-2$ 
nodes is connected to the current node
(otherwise, the only possible move would have 
been to hop back to the previous node).
This leaves $N-3$ nodes such that each one of them is connected to the
current node with probability $p$.
Thus, the expectation value of the number of neighbors of the current
node, to which the NBW may hop in the $t+1$ time step, 
is $(N-3)p+1$.
Due to the local tree-like structure of ER networks, 
it is extremely unlikely that
the one node which is guaranteed to be connected to the current node has
already been visited. 
This is due to the fact that the path from such earlier
visit of this adjacent node all the way to the current node is essentially a loop. 
Therefore, we
conclude that this adjacent node has not yet been visited.
Since the number of yet unvisited nodes is $N-t-1$, we conclude that
the current node is expected to have
$(N-t-2)p+1$ neighbors which have not yet been visited.
As a result, the probability that the RW will hop into one of the
yet-unvisited nodes is given by

\begin{equation}
P_{\rm ret}(d>t | d> t-1) = \frac{(N-t-2)p+1}{(N-3)p+1}.
\label{eq:P_r00}
\end{equation}

\noindent
Inserting 
$c=(N-1)p$
and 
$c(t)=(N-t-1)p$
we obtain

\begin{equation} 
P_{\rm ret}(d>t|d>t-1) = \frac{c(t)-p+1}{c-2p+1}. 
\label{eq:P_r2}
\end{equation}

\noindent
In the asymptotic limit this expression can be
approximated by

\begin{equation} 
P_{\rm ret}(d>t|d>t-1) = \frac{c(t)+1}{c+1}. 
\label{eq:P_r}
\end{equation}

\noindent
Combining the results presented above, it is found that
the probability that the path of the NBW will not terminate 
at the $t+1$ time step
is given by the conditional probability

\begin{equation} 
P(d > t|d > t-1) = 
\left[ \frac{c(t)-p+1}{c-2p+1} \right]
\left(1 - e^{-c} \right).
\label{eq:cond3p}
\end{equation}

\begin{figure}
\centerline{
\includegraphics[width=7.5cm]{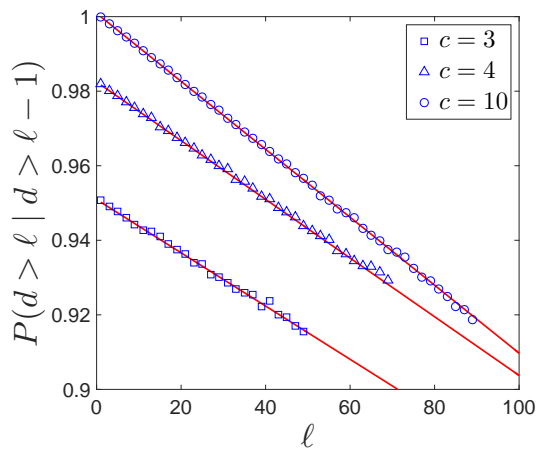}
}
\caption{
Theoretical results
(solid lines)
for the conditional probability 
$P(d > \ell|d > \ell-1)$ 
vs. $\ell$,
obtained from Eq.
(\ref{eq:cond3p}),
and the results obtained from computer simulations 
(symbols)
of NBWs on ER networks of
size $N=1000$ and mean degrees $c=3$, $4$ and $10$ 
(squares, triangles and circles, respectively).
The analytical and numerical results are found to
be in very good agreement.
} 
\label{fig:1}
\end{figure}

In Fig. 
\ref{fig:1} 
we present
the conditional probability 
$P(d > t|d > t-1)$ 
vs. $t$ for an NBW on an ER 
network 
of size $N=1000$ and three values of $c$. 
The analytical results 
(solid lines) 
obtained from Eq.
(\ref{eq:cond3p})
are found to be in very good agreement with numerical simulations
(symbols),
confirming the validity of this equation.
Note that the numerical results become more noisy as $t$
increases, due to diminishing statistics, 
and eventually terminate.
This is particularly apparent for the 
smaller values of $c$.

The probability that the path length of the NBW will
be longer than 
$\ell$ is given by

\begin{equation} 
P(d>\ell) = P(d>0) \prod_{t=1}^{\ell} P(d > t|d > t-1),
\label{eq:cond1}
\end{equation}

\noindent
where $P(d>0)=1$,
since the initial node is not isolated.
Using Eq.
(\ref{eq:cond3})
the probability 
$P(d>\ell)$ 
can be written as a product of the form

\begin{equation} 
P(d>\ell) =  P_{\rm ret}(d>\ell) P_{\rm trap}(d>\ell),
\label{eq:cond2}
\end{equation}

\noindent
where

\begin{equation} 
P_{\rm trap}(d>\ell) =  \prod_{t=1}^{\ell} 
\left( 1 - e^{-c} \right),
\label{eq:cond4}
\end{equation}

\noindent
and

\begin{equation} 
P_{\rm ret}(d>\ell) =  \prod_{t=1}^{\ell} 
\left[ \frac{c(t)-p+1}{c-2p+1} \right].
\label{eq:cond6}
\end{equation}

In Appendix A we evaluate the probabilities
$P_{\rm trap}(d>\ell)$,
given by Eq. (\ref{eq:cond4})
and
$P_{\rm ret}(d>\ell)$,
given by Eq. (\ref{eq:cond6}).
Combining the results for
$P_{\rm trap}(d>\ell)$
and for
$P_{\rm ret}(d>\ell)$
we obtain

\begin{eqnarray}
\ln [P(d>\ell)]
&\simeq&
 \left(\ell-\frac{1}{2}-\alpha^2 \right) 
\ln \left(1 - \frac{\ell-1/2}{\alpha^2}\right)-(\beta+1)\ell
\nonumber \\
&+&
\left(\alpha^2 - \frac{1}{2} \right) 
\ln \left(1 - \frac{1}{2\alpha^2}\right)
+1,
\label{eq:tail_tr}
\end{eqnarray}

\noindent
where

\begin{equation}
\alpha=\sqrt{\frac{N\left(c+1\right)}{c}},
\label{eq:alpha}
\end{equation}

\noindent
and

\begin{equation}
\beta = - \ln \left( 1-e^{-c} \right).
\label{eq:beta}
\end{equation}

\noindent
Assuming that the NBW paths are short 
compared to the network size, namely
$\ell \ll N$, 
one can use the expansion

\begin{equation}
\ln\left(1-\frac{x}{\alpha^2}\right)
\simeq
-\frac{x}{\alpha^2}-\frac{1}{2}
\left(\frac{x}{\alpha^2}\right)^{2}
+ O\left( \frac{x^3}{\alpha^6} \right).
\label{eq:lnapprox}
\end{equation}

\noindent
Plugging this approximation in Eq.
(\ref{eq:tail_tr})
yields

\begin{equation}
P\left(d>\ell\right)
\simeq
\exp \left[-\frac{\ell(\ell-1)}{2 \alpha^2} - \beta \ell \right].
\label{eq:P(d>ell)_simp}
\end{equation}

\noindent
Thus, the distribution of path lengths is a product of 
an exponential distribution and a
Rayleigh distribution,
which is a special case of the Weibull distribution
\cite{Papoulis2002}.
Considering the next order in the series expansion of Eq.
(\ref{eq:lnapprox})
we find that the relative error in 
Eq. (\ref{eq:P(d>ell)_simp})
for $P(d>\ell)$
due to the truncation of the Taylor expansion after the second order
is $\eta_{\rm TE} =  \ell^3/(6 \alpha^4)$,
which scales like $\ell^3/N^2$.
This error is very small as long as $\ell \ll N^{1/2}$.
Note that paths of length $\ell \simeq N^{1/2}$,
for which the error 
in $P(d>\ell)$
is noticeable, become prevalent only
in the limit of dense networks, where $c > N^{1/2}$.
The probability density function 
$P(d=\ell)$
can be obtained by 

\begin{equation}
P\left(d=\ell\right) = P\left(d>\ell-1\right)-P\left(d>\ell\right).
\label{eq:pdf}
\end{equation}

\begin{figure}
\centerline{
\includegraphics[width=13cm]{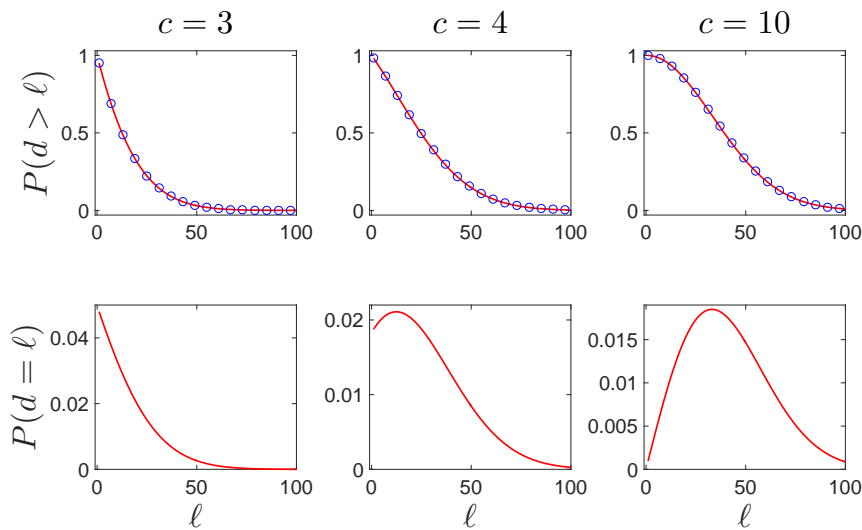}
}
\caption{
The tail distributions, 
$P(d > \ell)$,
of the first hitting times
of NBWs, 
vs. $\ell$,
for ER networks
of size $N=1000$ 
and 
$c=3$, $4$ and $10$.
The theoretical results 
(solid lines),
obtained from Eq.
(\ref{eq:tail_tr}) 
are found to be in excellent agreement 
with the results obtained
from numerical simulations (circles).
The corresponding probability density functions,
$P(d=\ell)$,
obtained from Eqs.  
(\ref{eq:tail_tr})
and
(\ref{eq:pdf}),
are shown in the bottom row. The agreement with the
numerical results is already established in the top 
row and therefore the numerical
data is not shown in the bottom row. 
}
\label{fig:2}
\end{figure}

In Fig.
\ref{fig:2}
we present the tail distributions
$P(d>\ell)$
vs.
$\ell$
of first hitting times of NBWs on ER networks of size
$N=1000$ and 
mean degrees 
$c=3$, $4$ and $10$ (top row).
The theoretical results (solid lines) 
were obtained from
Eq.
(\ref{eq:tail_tr}).
They are found to be in excellent agreement 
with the numerical simulations
(symbols).
The corresponding probability density functions,
$P(d=\ell)$,
obtained from Eqs. 
(\ref{eq:tail_tr}) 
and
(\ref{eq:pdf}),
are shown in the bottom row.

\section{Central and dispersion measures}

In order to characterize the distribution of 
first hitting times of NBWs on ER networks we
derive expressions for the mean and median of 
this distribution.
The mean of the distribution can be obtained
from the tail-sum formula

\begin{equation}
\ell_{\rm mean}(N,c) =
\sum_{\ell=0}^{N-2} P(d>\ell). 
\label{eq:tailsum1}
\end{equation}

\noindent
Assuming that the initial node is not isolated
and expressing the sum 
as an integral we obtain

\begin{equation}
\ell_{\rm mean}(N,c) \simeq
1 + \int_{1/2}^{N-3/2} P(d>\ell)d\ell,
\label{eq:ell_mean1}
\end{equation}

\noindent
where the limits of the integration are set
such that the summation over each integer, $i$, is replaced by an
integral over the range $(i-1/2,i+1/2)$.
Inserting 
$P(d>\ell)$
from Eq.
(\ref{eq:P(d>ell)_simp})
and solving the integral we obtain

\begin{equation}
\ell_{\rm mean}
=1+\sqrt{\frac{\pi}{2}}\alpha e^{\frac{(\alpha^{2}\beta - 1)\beta}{2}}
\left[{\rm erf}\left(\frac{\alpha^2\beta +N - 2}{\sqrt{2} \alpha}\right)
-{\rm erf}\left(\frac{\alpha\beta}{ \sqrt{2} }\right)\right].
\label{eq:ell_mean4}
\end{equation}

\noindent
where ${\rm erf}(x)$ is the error function, also called Gauss error function.
This function exhibits a sigmoid shape. For 
$|x| \ll 1$ 
it can be
approximated by 
${\rm erf}(x) \simeq 2x/\sqrt{\pi}$ 
while for 
$|x| > 1$ 
it quickly converges to 
${\rm erf}(x) \rightarrow {\rm sign}(x)$.
While the arguments of both ${\rm erf}$ functions 
in Eq.
(\ref{eq:ell_mean4})
are large,
the argument of the first ${\rm erf}$ function is much
larger than the argument of the second.
Therefore, one can safely set the first ${\rm erf}$ function to be equal to $1$,
and obtain

\begin{equation}
\ell_{\rm mean}
\simeq
1+\sqrt{ \frac{\pi}{2} }\alpha e^{\frac{(\alpha^{2}\beta - 1)\beta}{2}}
\left[1-{\rm erf} \left(\frac{\alpha\beta}{\sqrt{2}}\right)\right].
\label{eq:ell_mean5}
\end{equation}

\begin{figure}
\centerline{
\includegraphics[width=18cm]{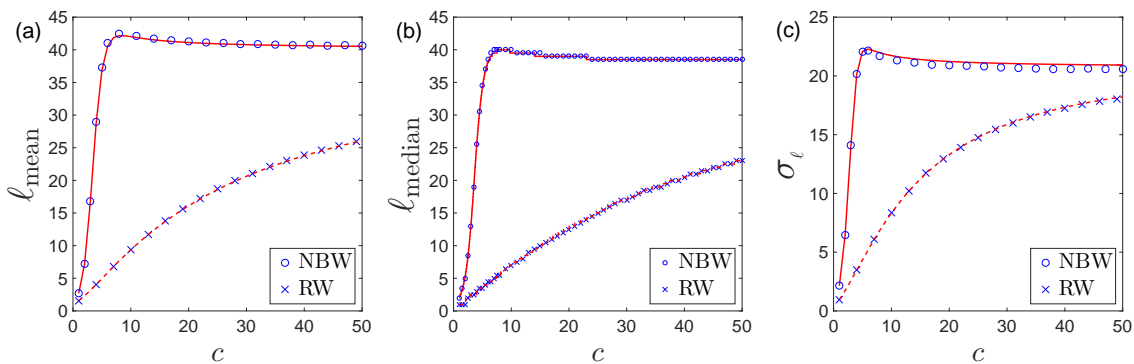}
}
\caption{
The mean 
$\ell_{mean}$ (a), 
the median
$\ell_{median}$ (b)
and the standard deviation 
$\sigma_{\ell}$ (c), 
of the distribution of first hitting times
of NBWs,
as a function of the mean degree,
$c$, for ER networks of size
$N=1000$.
The analytical results 
(solid lines), 
obtained from Eqs.
(\ref{eq:ell_mean5}),
(\ref{eq:ell_median1})
and
(\ref{eq:sigmaell}),
respectively,
are in excellent agreement with numerical 
simulations 
(circles).
For comparison, we also present analytical 
(dashed lines) 
and numerical ($\times$)
results for 
the mean,
$\ell_{mean}$ (a), 
the median,
$\ell_{median}$ (b)
and the standard deviation,
$\sigma_{\ell}$ (c),
of the distribution of first hitting times of 
RWs on the same network.
Unlike the NBWs, in which the three measures sharply increase
and then saturate as $c$ is increased,
in case of the RW they rise gradually. 
}
\label{fig:3}
\end{figure}

In Fig.
\ref{fig:3}(a)
we present the mean,
$\ell_{\rm mean}$,
of the distribution of first hitting times
of NBWs
as a function of the mean degree $c$,
for ER networks of size $N=1000$.
The agreement between the theoretical results,
obtained from Eq.
(\ref{eq:ell_mean5})
and the numerical simulations is very good for all values of
$c$.

To obtain a more complete characterization of the distribution of first
hitting times, it is also useful to evaluate its median, 
$\ell_{\rm median}$.
Here the
median is defined as the value of $\ell$ for which

\begin{equation}
| P(d>\ell) - P(d<\ell) | \rightarrow {\rm min},
\label{eq:ell_median1}
\end{equation}

\noindent
where $\ell$ may take either an integer or a half-integer value.
In Fig.
\ref{fig:3}(b)
we present the median, 
$\ell_{\rm median}$,
of the distribution of first hitting times
of NBWs
as a function of the mean degree $c$,
for ER networks of size $N=1000$.
The agreement between the theoretical results
and the numerical simulations is very good for all values of
$c$.

The moments of the distribution of NBW path lengths,
$\langle \ell^n \rangle$,
are given by
the tail-sum formula
\cite{Pitman1993}

\begin{equation}
\langle \ell^n \rangle = 
\sum_{\ell=0}^{N-2} [(\ell+1)^n - \ell^n] P(d>\ell).
\label{eq:tail_sumn}
\end{equation}

\noindent
Using this formula to evaluate the second moment 
and replacing the sum by an integral we obtain

\begin{equation}
\langle \ell^{2} \rangle = 1 +
\int_{\frac{1}{2}}^{N-\frac{3}{2}} \left(2\ell+1\right)
\exp\left[-\frac{\ell(\ell-1)}{2\alpha^2} - \beta\ell\right] d\ell.
\label{eq:ell2}
\end{equation}

\noindent
As in Eq. 
(\ref{eq:ell_mean4}),
the solution of this integral consists of two 
${\rm erf}$ 
functions.
For $N \gg 1$
one can approximate the upper limit of the first error function
by $1$, 
replace 
$N-3/2$ by $N$ 
and neglect the term 
$\exp(-N^2)$,
to obtain

\begin{equation}
\langle \ell^2 \rangle
\simeq
1 + 2 \alpha^2 
e^{- \frac{\beta}{2}}
+
\sqrt{2\pi}\alpha \left(1 - \alpha^{2} \beta \right)
e^{(\alpha^2 \beta - 1)\beta/2}  
\left[1 - {\rm erf} \left(\frac{\alpha\beta}{\sqrt{2}}\right)\right]. 
\label{eq:ell_sqr}
\end{equation}

\noindent
The standard deviation,
$\sigma_{\ell}(c)$,
of the distribution of path lengths is
thus given by

\begin{equation}
\sigma_{\ell}^2(c) = \langle \ell^2 \rangle - (\ell_{\rm mean})^{2},
\label{eq:sigmaell}
\end{equation}

\noindent
where $\langle \ell^2 \rangle$
is given by Eq.
(\ref{eq:ell_sqr})
and
$\ell_{mean}$
is given by Eq.
(\ref{eq:ell_mean5}).

In Fig.
\ref{fig:3}(c)
we present the standard deviation, 
$\sigma_{\ell}(c)$
of the distribution of first hitting times
of NBWs
as a function of the mean degree, $c$,
for ER networks of size $N=1000$.
The agreement between the theoretical results,
obtained from Eq. 
(\ref{eq:sigmaell}),
and the numerical simulations is very good for all values of
$c$.

\section{Analysis of the two termination mechanisms}

The NBW model studied here may terminate either by the
trapping scenario or by the retracing scenario.
The trapping mechanism may occur starting from the
second step of the NBW. 
The probability of trapping 
is 
$\exp(-c)$ 
at any time step afterwards, 
regardless of the number
of steps already pursued.
The termination by retracing takes place when the NBW steps into
a node which it has already visited before.
In this case, the path forms a loop
which starts at the first visit to the termination node and ends in
the second visit.
Termination by the retracing scenario may occur starting from
the third time step of the NBW. 
The probability that the NBW will terminate
due to retracing increases in time.
This is due to the fact that each visited node
becomes a potential termination site.
It is thus expected that paths that terminate after a small number of
steps are likely to terminate by trapping, while paths
which survive for a long time are more likely to terminate by
retracing. Below we present a detailed analysis of the probabilities
of an NBW to terminate by trapping or by retracing. 

Consider an NBW on an ER network, which starts from a random, non-isolated node,
hops to a new node at each of the first $\ell$ time steps,
and terminates
at the $\ell+1$ step.
Since the termination step is not counted as a part of the
path, the length of such NBW path is $d=\ell$.
The probability distribution function of the NBW path lengths,
$P(d=\ell)$, 
is given by Eq.
(\ref{eq:pdf}).
We denote by $p_{\rm trap}$
the probability that an NBW 
starting from a random initial node
will eventually terminate
by the trapping scenario
and by $p_{\rm ret}$ 
the probability that it will terminate
by the retracing scenario.
Since these are the only two termination mechanisms in the NBW model,
the two probabilities must satisfy
$p_{\rm trap} + p_{\rm ret} = 1$.

While the overall distribution of path lengths is given by
$P(d=\ell)$, 
one expects the distribution 
$P(d=\ell | {\rm trap})$, of 
paths terminated by trapping, 
to differ from the distribution
$P(d=\ell | {\rm ret})$, 
of paths terminated by retracing.
These conditional probability distributions 
are normalized,
namely they satisfy

\begin{equation}
\sum_{t=1}^{N-1} P(d= t | {\rm trap}) =1,
\label{eq:normb}
\end{equation}

\noindent
and

\begin{equation}
\sum_{t=2}^{N-1} P(d= t | {\rm ret}) =1.
\label{eq:normr}
\end{equation}

\noindent
The distribution of path lengths can be expressed in terms of 
the conditional distributions according to

\begin{equation}  
P(d=\ell) = p_{\rm trap} P(d=\ell | {\rm trap}) + p_{\rm ret} P(d=\ell | {\rm ret}).
\label{eq:br}
\end{equation}

\noindent
The first term on the right hand side of Eq.
(\ref{eq:br}) 
can be written as

\begin{equation}
p_{\rm trap} P(d=\ell | {\rm trap}) =
P(d>\ell-1) \left[1-P_{\rm trap}(d>\ell|d>\ell-1) \right],
\label{eq:p_b}
\end{equation}

\noindent
namely as the probability that the RW will pursue $\ell$ steps
and will terminate at the $\ell+1$ step by the trapping scenario.
The second term on the right hand side of Eq.
(\ref{eq:br}) 
can be written as

\begin{equation}
\fl
p_{\rm ret} P(d=\ell | r) =
P(d>\ell-1) P_{\rm trap}(d>\ell|d>\ell-1)
\left[1-P_{\rm ret}(d>\ell|d>\ell-1)\right],
\label{eq:p_r}
\end{equation}

\noindent
namely as the probability that the RW will pursue $\ell$ steps,
then in the $\ell+1$ step it will not get trapped but will
retrace its path by stepping into a node which was already visited at least two steps earlier.

Summing up both sides of Eq.
(\ref{eq:p_b})
over all integer values of $\ell$ we obtain

\begin{equation}
p_{\rm trap} = e^{-c}  \sum_{\ell=1}^{N-1} P(d>\ell-1).
\end{equation}

\noindent
Using the tail-sum formula, 
Eq. 
(\ref{eq:tailsum1}),
we find that the
probability that the NBW will terminate 
by the trapping scenario is

\begin{equation}
p_{\rm trap} = e^{-c} \ell_{\rm mean}.
\label{eq:p_b2}
\end{equation}

\noindent
Therefore, the probability of the NBW to terminate by
retracing its path is

\begin{equation}
p_{\rm ret} = 1 - e^{-c} 
\ell_{\rm mean}.
\label{eq:p_r2}
\end{equation}

\noindent
Using Eq. 
(\ref{eq:p_b})
the conditional probability
$P(d=\ell | {\rm trap})$ 
can be written in the form

\begin{equation}
P(d=\ell | {\rm trap}) = \frac{P(d>\ell-1)}{\ell_{\rm mean}},
\label{eq:p_lbeq}
\end{equation}

\noindent
where
$P(d>\ell-1)$
is given by Eq.
(\ref{eq:P(d>ell)_simp}).
Similarly, the conditional probability
$P(d=\ell | {\rm ret})$
takes the form

\begin{equation}
P(d=\ell | {\rm ret}) =  \left( \frac{ 1-e^{-c} }{c+1} \right) 
\left[ \frac{ c -c(\ell) }{1 - e^{-c} \ell_{\rm mean}} \right] P(d>\ell-1), 
\label{eq:p_lreq}
\end{equation}

\noindent
where $c(\ell)$ is given by Eq.
(\ref{eq:coft}).
The corresponding tail distributions
take the form

\begin{equation}
P(d > \ell | {\rm trap}) = 
\frac{\sum\limits_{t=\ell+1}^{N-1} P(d>t-1)}{\ell_{\rm mean}},
\label{eq:p_lbgt}
\end{equation}

\noindent
and

\begin{equation}
P(d > \ell | {\rm ret}) =  \left( \frac{1-e^{-c}}{c+1} \right) 
\sum\limits_{t=\ell+1}^{N-1} 
\left[ \frac{c - c(t)}{1-e^{-c}\ell_{\rm mean}} \right] P(d>t-1).
\label{eq:p_lrgt}
\end{equation}

Given that the path of an NBW terminated after $\ell$ steps, it is interesting
to evaluate the conditional probabilities 
$P({\rm trap} | d=\ell)$ 
and
$P({\rm ret} | d=\ell)$,
that the termination was caused by trapping or by
retracing, respectively.
Using Bayes' theorem,
these probabilities can be expressed by

\begin{equation}
P({\rm trap} | d=\ell) = \frac{p_{\rm trap} P(d=\ell | {\rm trap})}{P(d=\ell)}
\end{equation}

\noindent
and

\begin{equation}
P({\rm ret} | d=\ell) = \frac{p_{\rm ret} P(d=\ell | {\rm ret})}{P(d=\ell)}.
\end{equation}

\noindent
Clearly, these distributions satisfy
$P({\rm trap} | d=\ell) + P({\rm ret} | d=\ell) =1$.
Inserting the conditional probabilities
$P(d=\ell | {\rm trap})$
and
$P(d=\ell | {\rm ret})$
from Eqs.
(\ref{eq:p_lbeq})
and
(\ref{eq:p_lreq}),
respectively, we find that

\begin{equation}
P({\rm trap} | d=\ell) = 
\frac{P(d>\ell-1)}{e^{c} P(d=\ell)}
\label{eq:b_ell}
\end{equation}

\noindent
and

\begin{equation}
P({\rm ret} | d=\ell) = \left( 1-e^{-c} \right)
\frac{ \left[ c - c(\ell) \right]  }{(c+1)}
\frac{P(d>\ell-1)}{P(d=\ell)}.
\label{eq:b_ell2}
\end{equation}

\noindent
The corresponding tail distributions can be expressed in the form

\begin{equation}
P({\rm trap} | d>\ell) = 
\frac{\sum\limits_{t=\ell+1}^{N-1} 
P(d>t-1)}{ e^{c} P(d>\ell)}
\label{eq:b_ell3}
\end{equation}

\noindent
and

\begin{equation}
P({\rm ret} | d>\ell) = 
\left( \frac{1-e^{-c}}{c+1} \right) 
\sum_{t=\ell+1}^{N-1} 
\left[ c - c(t) \right] \frac{P(d>t-1)}{P(d>\ell)}.
\label{eq:r_ell}
\end{equation}

\noindent
These distributions also satisfy
$P({\rm trap} | d>\ell) + P({\rm ret} | d>\ell) =1$.

\begin{figure}
\centerline{
\includegraphics[width=9cm]{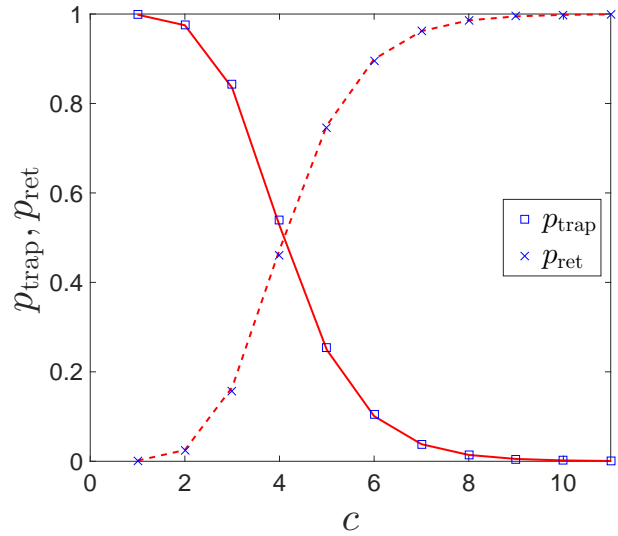}
}
\caption{
Analytical results for the
probabilities 
$p_{\rm trap}$ 
(solid line)
and 
$p_{\rm ret}$
(dashed line),
that an NBW on an ER network will terminate by  
trapping in a dead-end node
or by retracing of its path, 
respectively,
as a function of the mean degree, $c$.
The analytical results,
obtained from Eqs.
(\ref{eq:p_b2})
and
(\ref{eq:p_r2})
are found to be in excellent agreement with 
the results of numerical simulations 
($\square$ and $\times$, respectively).
}
\label{fig:4}
\end{figure}

In Fig.
\ref{fig:4}
we present the probability $p_{\rm trap}$ that an NBW will terminate due to trapping
and the probability $p_{\rm ret}$ that it will terminate due to retracing,
as a function
of the mean degree, $c$, for an ER network of size $N=1000$.
As expected, $p_{\rm trap}$ is a decreasing function of $c$ while $p_{\rm ret}$
is an increasing function.

\begin{figure}
\centerline{
\includegraphics[width=18cm]{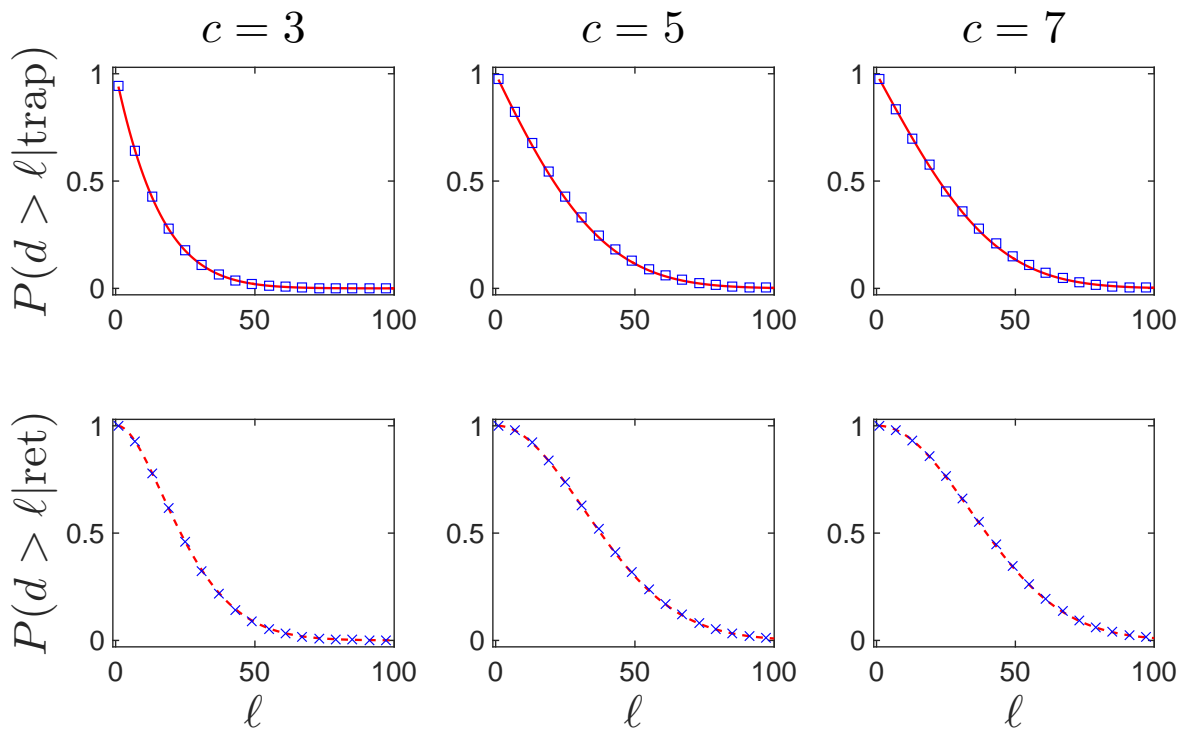}
}
\caption{
Analytical results for the
conditional tail distributions 
$P(d>\ell | {\rm trap})$
(solid lines)
and
$P(d>\ell | {\rm ret})$
(dashed lines)
of first hitting times 
vs. $\ell$,
for NBWs on an ER network, 
for paths terminated by trapping
(top row) or by retracing (bottom row),
respectively.
The results are shown for $N=1000$
and $c=3$, $5$ and $7$.
The theoretical results for 
$P(d>\ell | {\rm trap})$ 
are obtained from Eq.
(\ref{eq:p_lbgt}),
while the theoretical results for
$P(d>\ell | {\rm ret})$ 
are obtained from Eq.
(\ref{eq:p_lrgt}).
In both cases, the theoretical results are
found to be in excellent agreement with the numerical simulations
($\square$ and $\times$, respectively).
}
\label{fig:5}
\end{figure}

In Fig.
\ref{fig:5}
we present the probabilities
$P(d>\ell | {\rm trap})$
and
$P(d>\ell | {\rm ret})$
that an NBW path is of length larger than $\ell$, 
given that it terminated by trapping or by retracing,
respectively. The results are presented for $N=1000$ and
$c=3$, $5$ and $7$.
The analytical results (solid lines) are found to be in excellent agreement
with the numerical simulations (symbols).
In both cases, the paths tend to become longer as $c$ is increased.
However, for each value of $c$, the paths which terminate due to retracing
are typically longer than the paths which terminate due to trapping.

\begin{figure}
\centerline{
\includegraphics[width=18cm]{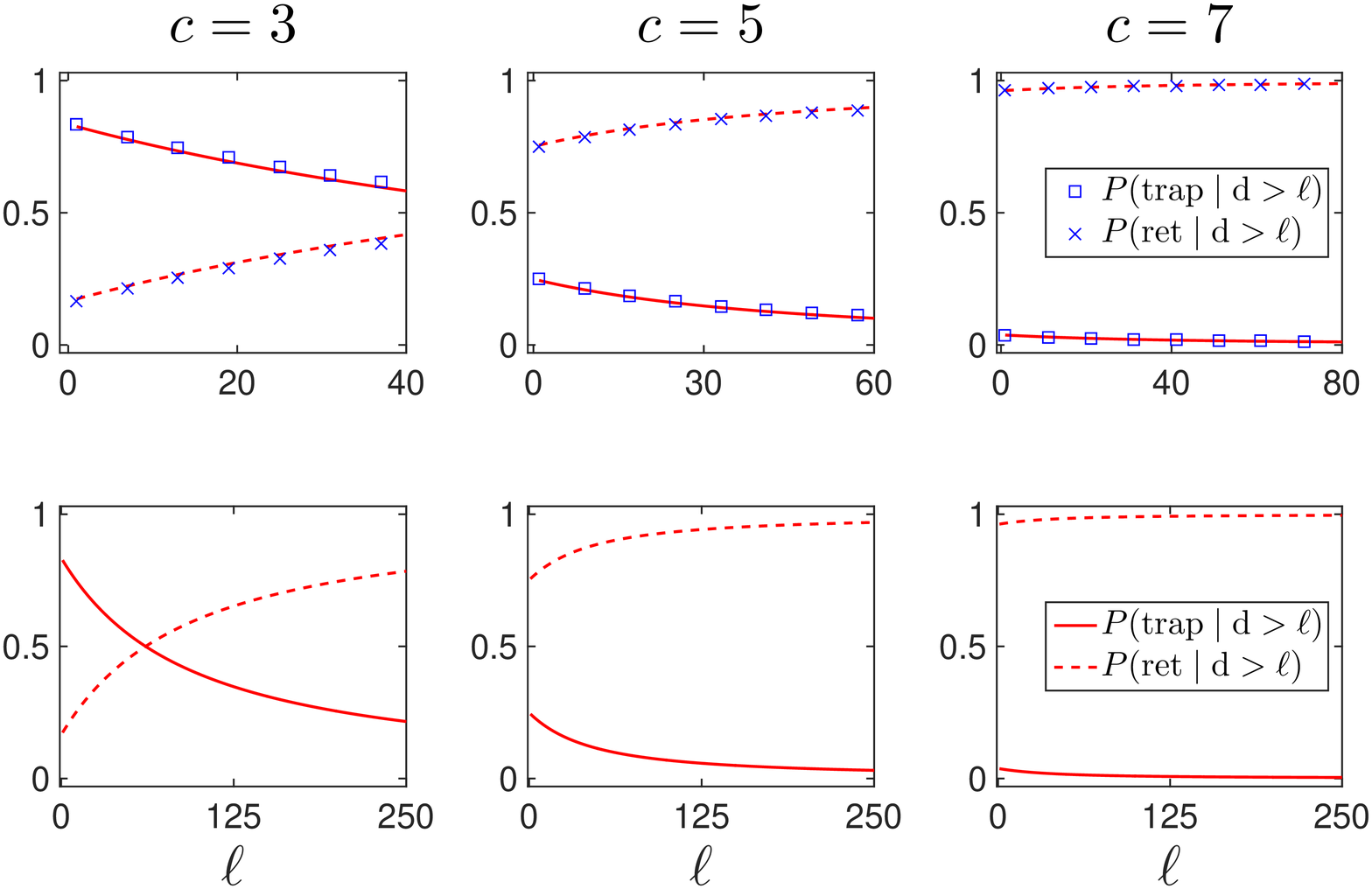}
}
\caption{
The conditional probabilities
$P({\rm trap} | d>\ell)$
and
$P({\rm ret} | d>\ell)$
that an NBW path will terminate by trapping or by retracing,
respectively, given that its length is larger than $\ell$,
are presented as a function of $\ell$.
Results are shown for an ER networks of size $N=1000$ and 
$c=3$, $5$ and $7$.
The theoretical results for
$P({\rm trap} | d>\ell)$ (solid lines)
are obtained from Eq.
(\ref{eq:b_ell3})
while the theoretical results for
$P( {\rm ret} | d>\ell)$ (dashed lines)
are obtained from Eq.
(\ref{eq:r_ell}).
In the top row these results are compared to the results of numerical
simulations (symbols) finding very good agreement. This comparison
is done for the range of path lengths which actually appear in the
numerical simulations and for which good statistics can be obtained.
Longer NBW paths which extend beyond this range become extremely
rare, so it is difficult to obtain sufficient numerical data.
However, in the bottom row we show the theoretical results
for a broader range of path lengths. 
It is found that 
$P({\rm trap} | d>\ell)$
is a monotonically decreasing function of $\ell$
while
$P({\rm ret} | d>\ell)$ is monotonically increasing.
}
\label{fig:6}
\end{figure}

In Fig. 
\ref{fig:6}
we present the probabilities
$P({\rm trap} | d>\ell)$
and
$P({\rm ret} | d>\ell)$
that an NBW will terminate due to trapping or retracing,
respectively.
Results are shown for ER networks of size $N=1000$ and 
$c=3$, $5$ and $7$.
The theoretical results for
$P({\rm trap} | d>\ell)$ (solid lines)
are obtained from Eq.
(\ref{eq:b_ell3})
while the theoretical results for
$P({\rm ret} | d>\ell)$ (dashed lines)
are obtained from Eq.
(\ref{eq:r_ell}).
As expected,
it is found that 
$P({\rm trap} | d>\ell)$
is a monotonically decreasing function of $\ell$
while
$P({\rm ret} | d>\ell)$ is monotonically increasing.
In the top row these results are compared to the results of numerical
simulations (symbols) finding very good agreement. 
This comparison
is done for the range of path lengths which actually appear in the
numerical simulations.
Longer NBW paths which extend beyond this range become extremely
rare, so it is difficult to obtain sufficient numerical data.
However, in the bottom row we show the theoretical results
for the entire range of path lengths. 

\begin{figure}
\centerline{
\includegraphics[width=9cm]{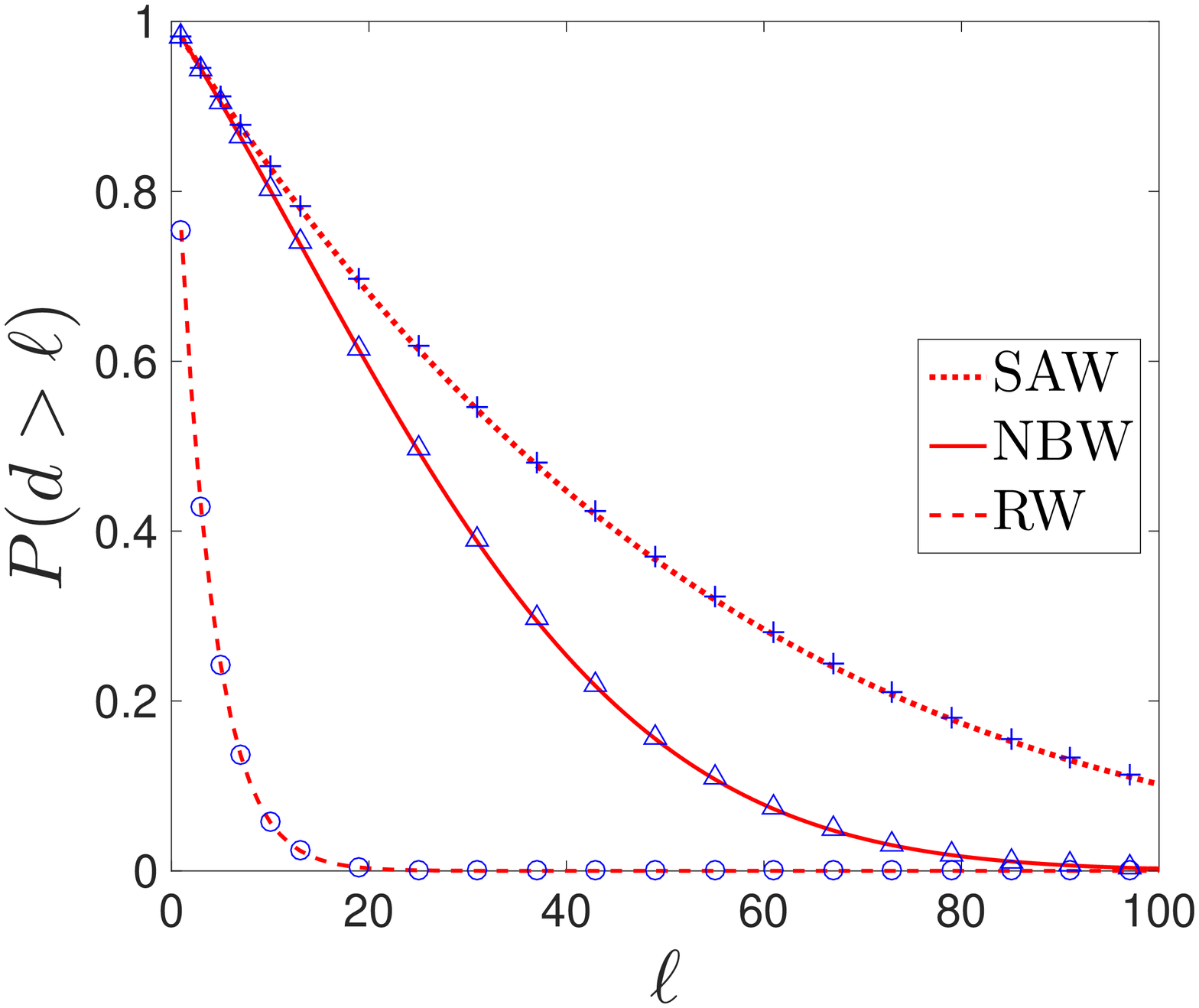}
}
\caption{
Theoretical results for the tail distribution
$P(d > \ell)$ 
of first hitting times of NBWs
(solid line),
the tail distribution of first hitting times 
of RWs (dashed line)
and the tail distribution
of last hitting times of SAWs
(dotted line),
on an ER network
of size $N=1000$ 
and 
$c=4$.
The results are in excellent agreement
with numerical simulations 
($\triangle$, $\circ$ and $+$, respectively).
The first hitting times of the NBWs are found to be
longer than those of the corresponding RWs,
but shorter than the last hitting times of the corresponding SAWs.
}
\label{fig:7}
\end{figure}

In Fig. 
\ref{fig:7}
we present 
the tail distribution
$P(d > \ell)$ 
of the first hitting times 
of the NBW (solid line)
on ER networks
of size $N=1000$ 
and 
$c=3$.
It is shown that the paths of the NBW
are much longer than those of the simple RW
(dashed line),
but much shorter than those of the SAW
(dotted line)
on the same network.

\section{Generalized non-backtracking random walk models}

We will now extend the analysis of the distribution of first
hitting times to generalized NBW models with $S>1$.
These NBWs keep track of the last $S$ nodes
they visited and avoid stepping into them again.
In the first $S$ time steps, the generalized NBW
behaves like an SAW model.
Hence, for $t \le S$
the conditional probability
$P_{\rm trap}(d>t|d>t-1)$
that the NBW will not be trapped at the $t+1$ time step
takes the form

\begin{equation}
P_{\rm trap}(d>t|d>t-1) =
1 - e^{-c(t)},
\label{eq:condSAW}
\end{equation}

\noindent
where $c(t)$ is given by Eq.
(\ref{eq:coft}),
while the conditional probability

\begin{equation}
P_{\rm ret}(d>t|d>t-1) = 1.
\label{eq:condSAWr}
\end{equation}

The conditional probability 
$P(d>t|d>t-1)$
can be expressed in the form

\begin{equation}
P(d>t|d>t-1) = 
P_{\rm trap}(d>t|d>t-1) 
P_{\rm ret}(d>t|d>t-1)
\label{eq:tailprod_tr}  
\end{equation}

\noindent
where
$P_{\rm trap}(d>t|d>t-1)$
is the probability that the path will not
terminate via the 
trapping scenario
and
$P_{\rm ret}(d>t|d>t-1)$
is the probability that it will not terminate
via the retracing scenario in the $t+1$ time step.

For $\ell \le S$ the probability that the NBW path length
is longer than $\ell$ is given by

\begin{equation} 
P(d>\ell) = 
P(d>0) 
\prod_{t=1}^{\ell} P_{\rm trap}(d > t|d > t-1)
\label{eq:tailSAW}
\end{equation}

\noindent
or by

\begin{equation}
P\left(d>\ell\right) = \prod_{t=1}^{\ell}
\left(1-e^{-c\left(t\right)}\right).
\end{equation}

In Ref. 
\cite{Tishby2016}
it was shown that the tail distribution of path lengths
of the SAW model takes the form

\begin{equation}
P_{\rm trap}(d>\ell)
\simeq
\exp \left\{\frac{N-1}{c}\left[{\rm Li}_{2}
\left(e^{- c + \frac{c}{2(N-1)}}
\right)-{\rm Li}_{2}\left(e^{-c + \frac{(2\ell+1)c}{2(N-1)} }
\right)\right]\right\}.
\label{eq:P_t(d>l)}
\end{equation}

\noindent
where 
${\rm Li}_{n}(x)$
is the Polylogarithm 
function
\cite{Olver2010}.
For large networks 
($N \gg 1$)
one can approximate
Eq.
(\ref{eq:P_t(d>l)})
by

\begin{equation}
P_{\rm trap}(d>\ell)
\simeq
\exp\left[- \eta
\left(e^{
b \ell}-1\right)\right],
\label{eq:tail2}
\end{equation}

\noindent
which is a discrete form of the 
Gompertz distribution
\cite{Gompertz1825,Johnson1995,Shklovskii2005,Ohishi2008},
where

\begin{equation}
b =\frac{c}{N}
\label{eq:scale}
\end{equation}

\noindent
is the scale parameter and

\begin{equation}
\eta = \frac{N}{c}e^{-c}
\label{eq:shape}
\end{equation}

\noindent
is the shape parameter.

Starting at time step $t=S+1$ 
the NBW may hop into nodes which were already
visited before at times 
$t^{\prime} = 0, 1, \dots, t-S-1$, 
thus causing termination of the path through the retracing mechanism.
The step in which the NBW hops into a 
previously visited node is not counted
as a part of the RW path. 
This means that the path length of an NBW which pursued
$\ell$ steps and was terminated in the $\ell+1$ step, 
is $d=\ell$. 
The path includes $\ell+1$ nodes, since the initial node is counted as
a part of the path.

In case that the NBW has already pursued $t>S$ steps 
without visiting any node more than once, 
the path length is guaranteed to be $d > t-1$.
At this point, the probability that the path will not
terminate by trapping in the $t+1$ step is 
denoted by the conditional probability
$P_{\rm trap}(d>t|d>t-1)$. 
We will now evaluate this probability for $t>S$.
The probability that the node
entered by the NBW at time $t$ 
does not have any other neighbor except for those 
nodes visited
in the last $S$ time steps is given by $e^{-c(S-1)}$.
This conditional probability is thus given by

\begin{equation}
P_{\rm trap}(d>t|d>t-1) = 1 - e^{-c(S-1)},
\label{eq:trap}
\end{equation}

\noindent
where
$c(S-1)$ is given by
Eq.
(\ref{eq:coft}).
For $t>S$ there are $N-S-1$ nodes, which may be 
connected to the current node with probability $p$.
Having eliminated the possibility of trapping
at the $t+1$ time step, 
guarantees that at least one of these 
nodes is actually connected to the current node.
Each one of the remaining 
$N-S-2$ 
nodes is connected to the current node with probability $p$.
Thus, the expectation value of the number of neighbors of the current
node to which the NBW may hop is
$(N-S-2)p+1$.
Due to the local tree-like structure of ER networks, 
it is extremely unlikely that
the one node which is guaranteed to be connected to the current node has
already been visited. 
This is due to the fact that the path from such earlier
visit all the way to the current node is essentially a loop. 
Therefore, we
conclude that this adjacent node has not been visited yet.
Since the number of yet unvisited nodes is $N-t-1$, 
the current node is expected to have
$(N-t-2)p+1$ 
neighbors which have not yet been visited.
As a result, 
the probability that in the $t+1$ time
step the NBW will hop into one of the
yet-unvisited nodes is given by

\begin{equation}
P_{\rm ret}(d>t | d> t-1) = \frac{(N-t-2)p+1}{(N-S-2)p+1}.
\label{eq:P_r01}
\end{equation}

\noindent
Using the fact that
$c=(N-1)p$
and 
$c(t)=(N-t-1)p$
we obtain

\begin{equation} 
P_{\rm ret}(d>t|d>t-1) = \frac{c(t)-p+1}{c-(S+1)p+1}. 
\label{eq:P_r2p}
\end{equation}

Summarizing the results so far,
we obtain that for $t \le S$
the conditional probability 
$P(d>t|d>t-1)$  
is given by
Eq.
(\ref{eq:tailprod_tr}),
while 
for $t > S$

\begin{equation}
P(d>t|d>t-1) = 
\left[ \frac{c(t)-p+1}{c-(S+1)p+1} \right]
\left[ 1 - e^{-c(S-1)} \right]. 
\label{eq:tr}
\end{equation}

\begin{figure}
\centerline{
\includegraphics[width=18cm]{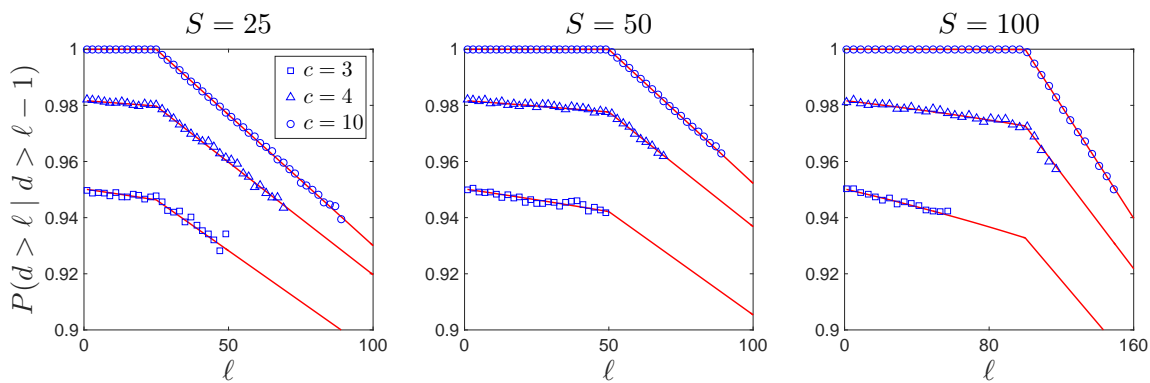}
}
\caption{
The conditional probability 
$P(d > \ell|d > \ell-1)$ 
vs. $\ell$,
obtained from Eqs.
(\ref{eq:condSAW})
and
(\ref{eq:tr})
(solid lines)
and 
from numerical simulations 
(symbols)
of generalized NBWs
with $S=25$ (left), $S=50$ (center) and $S=100$ (right),
on ER networks of
size $N=1000$ and mean degrees $c=3$, $4$ and $10$ 
(squares, triangles and circles, respectively).
The analytical and numerical results are found to
be in good agreement.
} 
\label{fig:8}
\end{figure}

\noindent
In Fig. 
\ref{fig:8} 
we present
the conditional probability 
$P(d > t|d > t-1)$ 
vs. $t$ for 
the generalized NBW model
with $S=25$, $50$ and $100$ on
an ER network 
of size $N=1000$ and mean degree $c=3$, $4$ and $10$. 
The analytical results (solid lines) obtained from Eqs.
(\ref{eq:condSAW})
and
(\ref{eq:tr})
are found to be in good agreement with numerical simulations
(symbols),
confirming the validity of these equations.
Note that the numerical results become more noisy as $t$
increases, due to diminishing statistics, 
and eventually terminate.
This is particularly apparent for the smaller values of $c$.

\begin{figure}
\centerline{
\includegraphics[width=18cm]{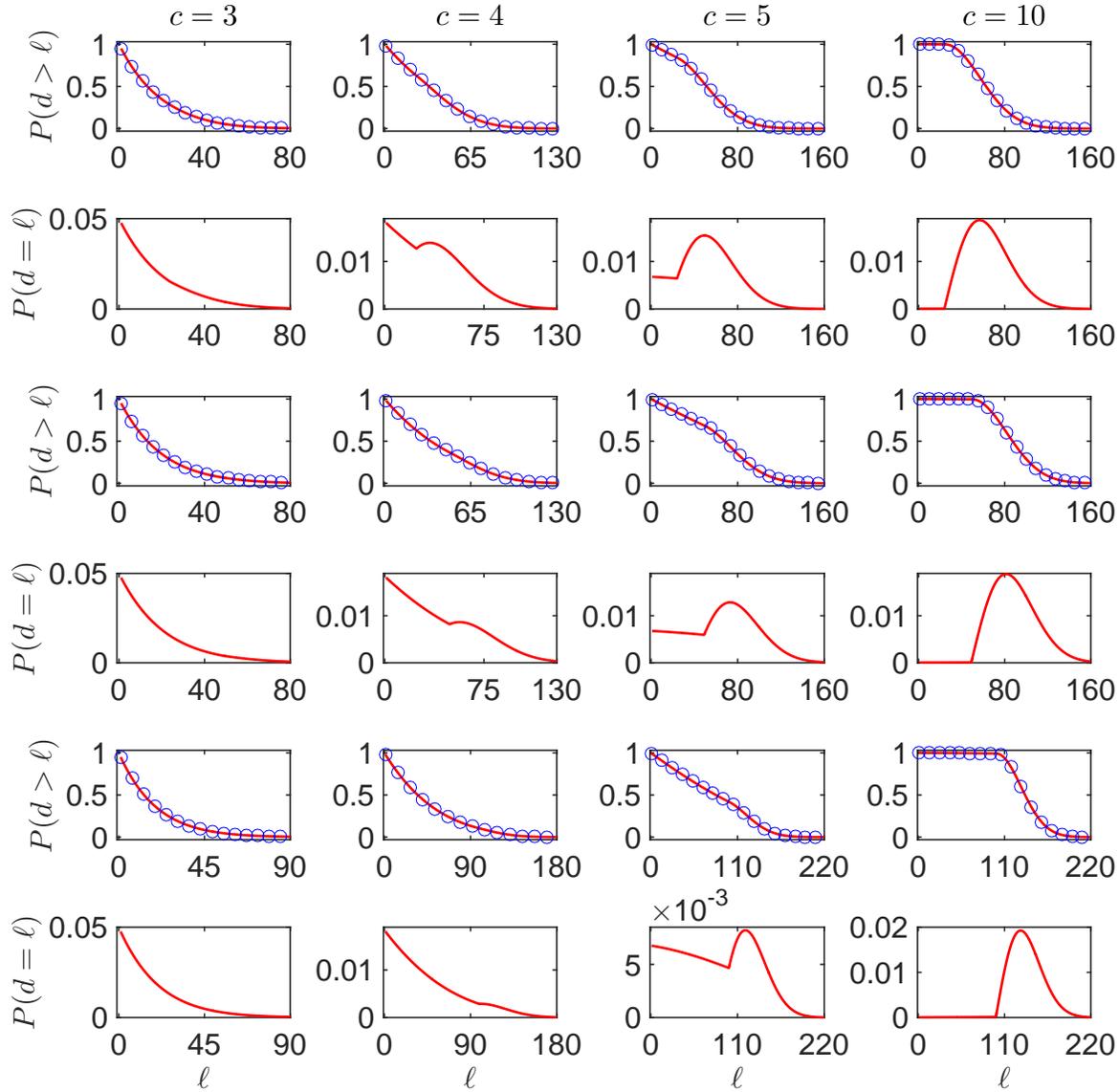} 
}
\caption{
The tail distributions,
$P(d > \ell)$,
of first hitting times
and the
corresponding probability density functions,
$P(d=\ell)$,
vs. $\ell$,
of generalized NBWs 
with $S=25$ (rows 1 and 2), $S=50$ (rows 3 and 4) 
and $S=100$ (rows 5 and 6)
on ER networks
of size $N=1000$ 
and 
$c=3$, $4$, $5$ and $10$.
The theoretical results, obtained from Eqs.
(\ref{eq:PS(d>ell)}) 
(solid lines),
are found to be in excellent agreement with the results obtained
from numerical simulations (circles).
The agreement with the
numerical results is already established for the tail
distributions and therefore the numerical
data is not shown for the probability density functions.  
}
\label{fig:9}
\end{figure}

In Appendix B we calculate the probabilities
$P_{\rm ret}(d>\ell)$
and
$P_{\rm trap}(d>\ell)$
and obtain

\begin{equation}
{P\left(d>\ell\right)=
\left\{
\begin{array}{ll}
\exp\left[-\eta \left(e^{b \ell}-1\right)\right]
& \ell \le S\\
\exp\left[-\eta \left(e^{b S}-1\right)\right]
\exp\left[-\frac{\ell_{S}(\ell_S - 1)}{2 \alpha^2} - \beta_S \ell_{S}
\right] & \ell\ge S+1,
\end{array}
\right.
}
\label{eq:PS(d>ell)}
\end{equation}

\noindent
where

\begin{equation}
\beta_{S}=-\ln\left[1-e^{- \left( 1 - \frac{S-1}{N-1} \right) c }\right],
\label{eq:beta_S}
\end{equation}

\noindent
and 
$\ell_S = \ell - (S-1)$.
In Fig.
\ref{fig:9}
we present the tail distributions of first hitting times
$P(d>\ell)$ (top rows)
vs.
$\ell$
for generalized NBWs with $S=25$, $50$ and $100$
on ER networks of size
$N=1000$ and 
mean degrees 
$c=3$, $4$, $5$ and $10$.
The theoretical results (solid lines) were obtained from
Eq.
(\ref{eq:PS(d>ell)}).
They are found to be in excellent agreement with the numerical simulations
(symbols).
The theoretical results for the probability distribution functions
are presented in the bottom rows.

In order to obtain an expression for the mean distance,
$\ell_{\rm mean}$, we use
the tail-sum formula
presented in Eq.
(\ref{eq:tail_sumn})
for $n=1$. 
Separating the sum 
into two parts, one for $\ell \le S$ and the other for $\ell > S$,
we obtain

\begin{equation}
\ell_{\rm mean} = 
\sum_{\ell=0}^{S}P_{\rm trap}\left(d>\ell\right) + \sum_{\ell=S+1}^{N-2}P\left(d>\ell\right).
\end{equation}

\noindent
In Appendix C we evaluate these sums and obtain

\begin{eqnarray}
\ell_{\rm mean} &\simeq&
1+ \frac{e^{\eta}}{b}
\left[{\rm Ei}\left( -\eta e^{b(S+{1}/{2})} \right)
-{\rm Ei}\left(-\eta e^{ \frac{b}{2}} \right)\right] 
\nonumber \\
&+&
\sqrt{\frac{\pi}{2}} \alpha
\exp\left[\frac{(\alpha^2 \beta_{S} - 1)\beta_S}{2} 
- \eta \left(e^{b S}-1\right)\right]
\nonumber \\
&\times& 
\left[{\rm erf}\left(\frac{\alpha^2\beta_{S}+(N-S)-2}{\sqrt{2}\alpha}\right)
-{\rm erf}\left(\frac{\alpha \beta_{S}}{\sqrt{2}}\right)\right],
\label{eq:ell_meanSfull}
\end{eqnarray}

\noindent
where the shape parameter 
$\eta$ is given by Eq. 
(\ref{eq:shape}),
the scale parameter $b$ is given by Eq.
(\ref{eq:scale}),
the parameter $\beta_S$ is given by
Eq.
(\ref{eq:beta_S})
and $\alpha$ is given by
(\ref{eq:alpha}).

\begin{figure}
\centerline{
\includegraphics[width=8cm]{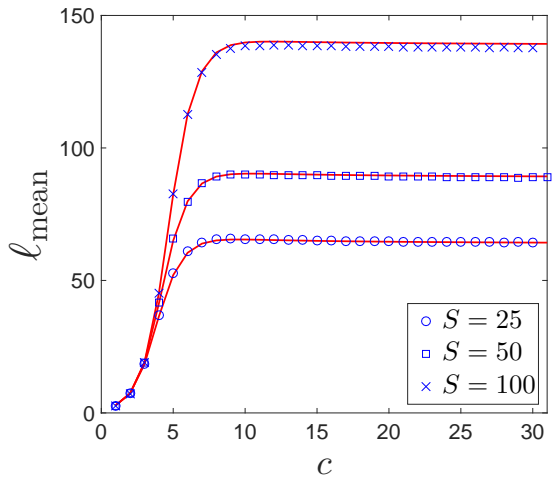}
}
\caption{
The mean of the distribution of first hitting times,
$\ell_{mean}$, as a function of the mean degree,
$c$, 
of generalized NBWs 
with $S=25$ ($\circ$), $S=50$ ($\square$) and $S=100$ ($\times$)
on ER networks of size
$N=1000$.
The analytical results (solid lines), 
obtained from Eq.
(\ref{eq:ell_meanSfull})
are in excellent agreement with numerical 
simulations 
(symbols). 
}
\label{fig:10}
\end{figure}

In Fig.
\ref{fig:10}
we present
the mean of the distribution of first hitting times,
$\ell_{mean}$, 
as a function of the mean degree,
$c$, 
of generalized NBWs
with $S=25$ (a), $S=50$ (b) and $S=100$ (c)
on ER networks of size
$N=1000$.
The analytical results (solid line), 
obtained from Eq.
(\ref{eq:ell_meanSfull})
are in excellent agreement with numerical 
simulations 
(circles). 
It is shown that 
$\ell_{mean}$
increases sharply as a function of $c$ 
and quickly saturates.
As $S$ is increased, the
saturation level of
$\ell_{mean}$
increases.

\begin{figure}
\centerline{
\includegraphics[width=8cm]{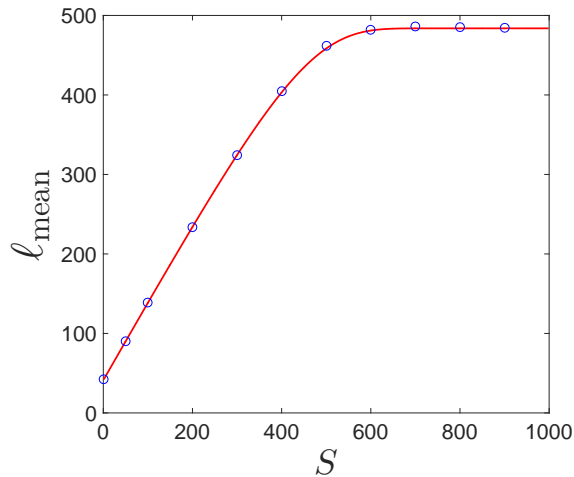}
}
\caption{
The mean of the distribution of first hitting times,
$\ell_{mean}$, as a function of $S$ for
generalized NBWs 
on an ER network of size
$N=1000$
and $c=10$ .
The analytical results (solid line), 
obtained from Eq.
(\ref{eq:ell_meanSfull}),
are in excellent agreement with numerical 
simulations 
(circles). 
}
\label{fig:11}
\end{figure}

In Fig.
\ref{fig:11} we present the
mean of the distribution of first hitting times,
$\ell_{mean}$, 
as a function of $S$ for
generalized NBWs 
on an ER network of size
$N=1000$
and $c=10$ .
The analytical results (solid line), 
obtained from Eq.
(\ref{eq:ell_meanSfull})
are in excellent agreement with numerical 
simulations 
(circles). 

The paths of generalized NBWs may terminate either by the trapping scenario or by
the retracing scenario. Therefore, the tail distribution 
$P(d>\ell)$ can be decomposed into two distributions,
each accounting for the paths terminated by one of the two scenarios.
First, we calculate the probabilities,
$p_{\rm trap}$ and $p_{\rm ret}$,
that a generalized NBW path will terminate by the
trapping scenario or by the retracing scenario, respectively.
The probability of termination by trapping, $p_{\rm trap}$ is given by

\begin{equation}
p_{\rm trap} = p_{\rm trap}[d \le S] + p_{\rm trap}[d \ge S+1],
\end{equation}

\noindent
where $p_{\rm trap}[d \le S]$ is the probability that a generalized NBW path
will terminate by trapping during the first $S$ time steps, while
$p_{\rm trap}[d>S]$ is the probability that it will terminate by trapping
after more than $S$ steps.
These probabilities are given by

\begin{equation}
p_{\rm trap}[d \le S] =
\sum_{\ell=1}^{S} 
P_{\rm trap}(d>\ell-1) [1-P_{\rm trap}(d>\ell | d>\ell - 1)],
\label{eq:p_tleS}
\end{equation}

\noindent
and

\begin{equation}
p_{\rm trap}[d \ge S+1] =
\sum_{\ell=S+1}^{N-2} 
P(d>\ell-1 | d \ge S+1) [1-P_{\rm trap}(d>\ell | d>\ell - 1)].
\label{eq:p_tgtS}
\end{equation}

\noindent
Summing up the terms on the right hand side of Eq.
(\ref{eq:p_tleS}),
it can be written in the form

\begin{equation}
p_{\rm trap}[d \le S] =
P_{\rm trap}(d>0) - P_{\rm trap}(d>S).
\label{eq:p_tleS123}
\end{equation}

\noindent
One can evaluate the sum on the right hand side of Eq.
(\ref{eq:p_tleS})
by converting it to an integral of the form

\begin{equation}
p_{\rm trap}[d \le S]
=\int_{1/2}^{S+1/2}
b \eta 
\exp\left[-\eta \left(e^{b (\ell-1)}-1\right)
+\frac{c}{N}\ell\right] 
d\ell.
\end{equation}

\noindent
Solving the integral we obtain

\begin{equation}
p_{\rm trap}[d \le S]
=\exp\left[\eta \left(1-e^{\frac{-b}{2}}\right)\right]
-\exp\left[\eta
\left(1-e^{b\left(S-\frac{1}{2}\right)}\right)\right]
\label{eq:b_S0}
\end{equation}

\noindent
Converting the sum on the right hand side of Eq.
(\ref{eq:p_tgtS})
to an integral and performing the integration we obtain

\begin{eqnarray}
\fl
p_{\rm trap}[d \ge S+1]
&=&
\left\{ 1  + \sqrt{ \frac{\pi}{2}}\alpha e^{\frac{(\alpha^{2}\beta_S -1)\beta_S}{2}}
\left[{\rm erf}\left(\frac{ \alpha^2 \beta_S+N-S-2}{\sqrt{2}\alpha}\right)
\right. \right.
\nonumber \\
\fl
&-& 
\left. \left.
{\rm erf}\left(\frac{ \alpha \beta_S }{ \sqrt{2} }\right)\right]
\right\} e^{-c\left(S-1\right)}.
\label{eq:b_S1}
\end{eqnarray}

\noindent
Combining the results of Eqs.
(\ref{eq:b_S0})
and
(\ref{eq:b_S1})
we find that

\begin{eqnarray}
\fl
&p_{\rm trap}
=\exp\left[\eta \left(1-e^{\frac{-b}{2}}\right)\right]
-\exp\left[\eta
\left(1-e^{b\left(S-\frac{1}{2}\right)}\right)\right]
\nonumber \\
\fl
&+
\left\{ 1+\sqrt{ \frac{\pi}{2}}\alpha e^{\frac{( \alpha^{2}\beta_S - 1)\beta_S}{2}}
\left[{\rm erf}\left(\frac{ \alpha^2 \beta_S+N-S - 2}{\sqrt{2}\alpha}\right)
-{\rm erf}\left(\frac{ \alpha \beta_S}{ \sqrt{2} }\right)\right]
\right\} e^{-c\left(S-1\right)}. 
\label{eq:p_trapping}
\end{eqnarray}

\noindent
The probability of termination by the retracing scenario is obtained from
$p_{\rm ret}=1-p_{\rm trap}$,
or

\begin{eqnarray}
\fl
&p_{\rm ret}
=1 - \exp\left[\eta \left(1-e^{\frac{-b}{2}}\right)\right]
+\exp\left[\eta
\left(1-e^{b\left(S-\frac{1}{2}\right)}\right)\right]
\nonumber \\
\fl
&-
\left\{ 1+\sqrt{ \frac{\pi}{2}}\alpha e^{\frac{(\alpha^{2}\beta_S - 1)\beta_S}{2}}
\left[{\rm erf}\left(\frac{ \alpha^2 \beta_S+N-S-2}{\sqrt{2}\alpha}\right)
-{\rm erf}\left(\frac{ \alpha \beta_S}{ \sqrt{2} }\right)\right]
\right\} e^{-c\left(S-1\right)}. 
\label{eq:p_trapping2}
\end{eqnarray}

\noindent
It turns out that for values of $c$ 
which are not too small,
the approximation involves in converting the
sum of Eq.
(\ref{eq:p_tgtS})
to an integral is very good and
Eqs. 
(\ref{eq:p_trapping})
and
(\ref{eq:p_trapping2})
provide accurate results for
$p_{\rm trap}$
and 
$p_{\rm ret}$,
respectively.
However, for small values of $c$, more accurate results are
obtained by using the recursion equations directly to evaluate
$p_{\rm trap}[d \le S]$
from Eq.
(\ref{eq:p_tleS123}).

\begin{figure}
\centerline{
\includegraphics[width=18cm]{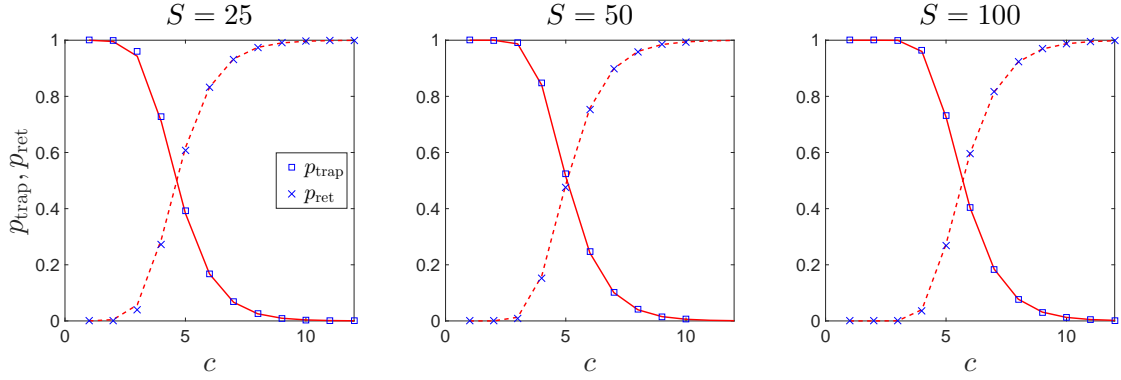}
}
\caption{
Analytical results for the
probabilities 
$p_{\rm trap}$ (solid lines) 
and 
$p_{\rm ret}$ (dashed lines)
that generalized NBWs
with $S=25$, $50$ and $100$
on an ER network will terminate by 
retracing their paths, 
or by being trapped in a dead-end node,
respectively,
as a function of the mean degree, $c$.
The theoretical results, obtained from Eqs.
(\ref{eq:p_trapping})
and
(\ref{eq:p_trapping2})
are found to be in excellent agreement with 
the results of numerical simulations 
(symbols).
}
\label{fig:12}
\end{figure}

In Fig.
\ref{fig:12}
we present the
probabilities 
$p_{\rm trap}$ 
and 
$p_{\rm ret}$
that generalized NBWs 
with $S=25$, $50$ and $100$
on an ER network will terminate via  
retracing of its path, 
or by being trapped in a dead-end node,
respectively,
as a function of the mean degree, $c$.
The theoretical results, obtained from Eqs.
(\ref{eq:p_trapping})
and
(\ref{eq:p_trapping2})
are found to be in excellent agreement with 
the results of numerical simulations 
(symbols).

\begin{figure}
\centerline{
\includegraphics[width=8cm]{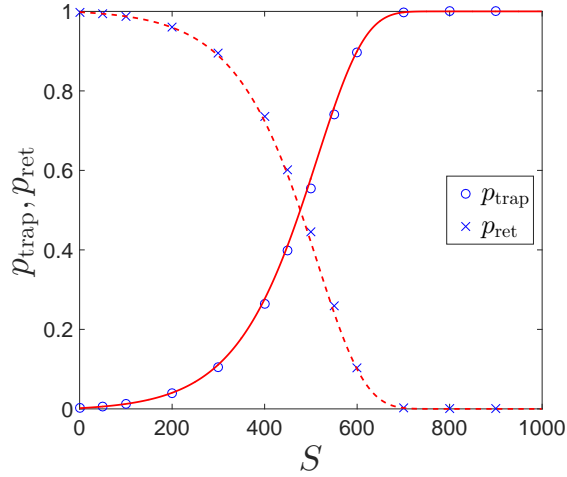}
}
\caption{
Analytical results for the probabilities 
$p_{\rm trap}$ (solid line)
and 
$p_{\rm ret}$ (dashed line) 
that a generalized NBW will terminate 
by trapping in a dead-end node or by  
retracing of its path, 
respectively,
as a function of the parameter $S$
on an ER network of size $N=1000$ and
$c=10$.
The theoretical results
for 
$p_{\rm trap}$ (solid line),
obtained from Eq.
(\ref{eq:p_trapping}),
and for
$p_{\rm ret}$ (dashed line),
obtained from Eq.
(\ref{eq:p_trapping2}),
are found to be in excellent agreement with 
the results of numerical simulations 
(symbols).
}
\label{fig:13}
\end{figure}

In Fig.
\ref{fig:13} 
we present the
probabilities 
$p_{\rm trap}$ 
and 
$p_{\rm ret}$
that a generalized NBW  path will terminate by
trapping or by retracing, respectively,
as a function of the parameter $S$
on an ER network of size $N=1000$ and
$c=10$.
The theoretical results, obtained from Eqs.
(\ref{eq:p_trapping})
and
(\ref{eq:p_trapping2})
are found to be in excellent agreement with 
the results of numerical simulations 
(symbols).

We will now calculate the conditional tail distribution of the path lengths of 
generalized NBW paths. The probability that an NBW path length will be
larger than $\ell$, given that it terminated by trapping is given by
$P\left(d>\ell| {\rm trap}\right)$,
while the probability that it will be larger than $\ell$ given that it
terminated by retracing is given by
$P\left(d>\ell| {\rm ret}\right)$.
The probability
$P\left(d>\ell| {\rm trap}\right)$
is given by

\begin{equation}
\fl
P\left(d>\ell| {\rm trap}\right)
=\frac{1}{p_{\rm trap}}\sum_{t=\ell+1}^{N-2}
P\left(d>t-1\right) \left[1-P_{\rm trap}\left(d>t|d>t-1\right) \right].
\end{equation}

\noindent
It can be expressed in the form

\begin{equation}
\fl
{ P\left(d>\ell| {\rm trap}\right)
=
\left\{
\begin{array}{ll}
\sum\limits_{t=\ell+1}^{S} 
e^{-c\left(t\right)}
\frac{P\left(d>t-1\right)}{p_{\rm trap}} 
+
{e^{-c\left(S-1\right)} }
\sum\limits_{t=S+1}^{N-1}
\frac{P\left(d>t-1\right)}{p_{\rm trap}}
&  \ell \le S
\\
{e^{-c\left(S-1\right)} }
\sum\limits_{t=\ell}^{N-1}
\frac{P\left(d>t-1\right)}{p_{\rm trap}}
&
\ell \ge S+1
\end{array} 
\right.
}
\label{eq:CondTailTrap}
\end{equation}

\noindent
The conditional tail distribution
$P\left(d>\ell| {\rm ret}\right)$
can be written in the form

\begin{eqnarray}
\fl
P\left(d>\ell| {\rm ret}\right)
&=&
\frac{1}{p_{\rm ret}}\sum\limits_{t=\ell}^{N-2}
P\left(d>t-1\right)  P_{\rm trap}\left(d>t|d>t-1\right)
\nonumber \\
&\times&
\left[1-P_{\rm ret}\left(d>t|d>t-1\right)\right].
\label{eq:Pdellret1}
\end{eqnarray}

\noindent
Thus, we find that 

\begin{equation}
\fl
{ P\left(d>\ell| {\rm ret}\right)
=
\left\{
\begin{array}{ll}
1
& \ell \le S   
\\
\frac{1-e^{-c\left(S-1\right)}}{p_{\rm ret}}
\sum\limits_{t=\ell}^{N-2}P\left(d>t-1\right)
\left[\frac{c-c\left(t\right)-Sp}{c-(S+1)p+1}\right]
& \ell \ge S+1
\end{array}
\right.
}
\label{eq:CondTailRet}
\end{equation}

\begin{figure}
\centerline{
\includegraphics[width=18cm]{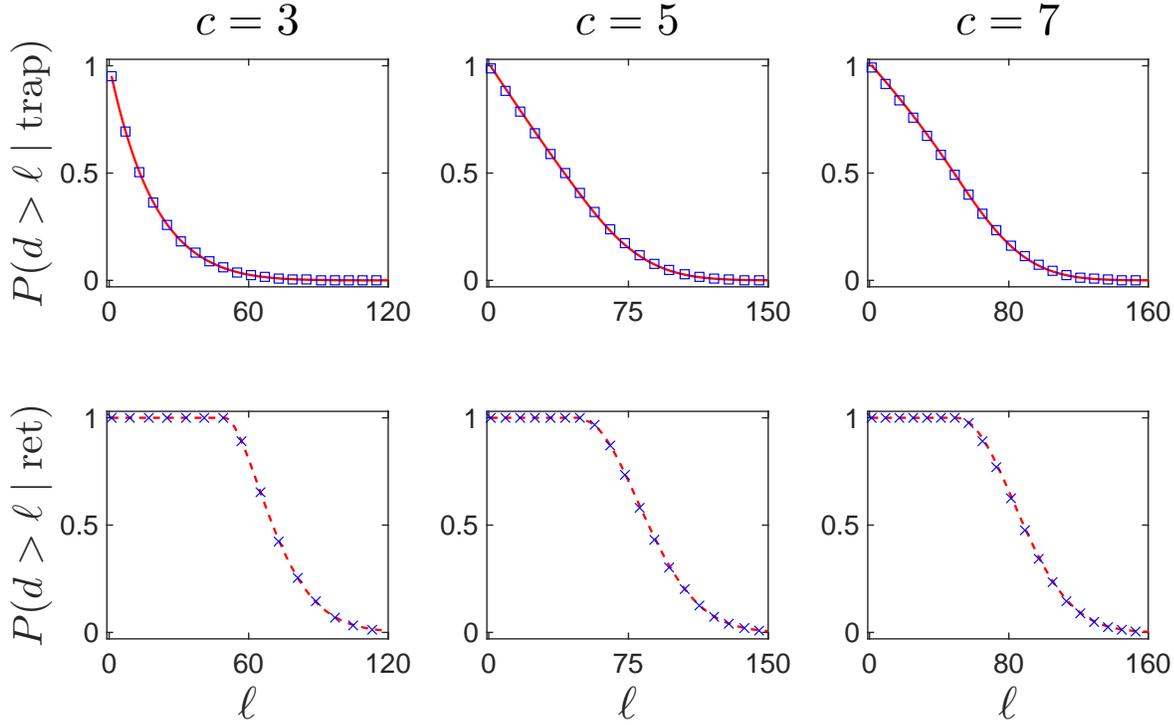}
}
\caption{
The conditional tail distributions 
$P(d>\ell | {\rm trap})$
and
$P(d>\ell | {\rm ret})$
of first hitting times 
vs. $\ell$,
of generalized NBWs with $S=50$, 
for paths terminated by trapping (top row) 
or by retracing (bottom row),
respectively.
The results are shown for 
ER networks of size
$N=1000$
and $c=3$, $5$ and $7$.
The theoretical results for 
$P(d>\ell | {\rm trap})$ 
are obtained from Eq.
(\ref{eq:CondTailTrap}),
while the theoretical results for
$P(d>\ell | {\rm ret})$ 
are obtained from Eq.
(\ref{eq:CondTailRet}).
In both cases, the theoretical results 
(solid and dashed lines lines, respectively) 
are found to be in excellent agreement 
with the numerical simulations
(symbols).
}
\label{fig:14}
\end{figure}

\noindent
In Fig.
\ref{fig:14}
we present the
conditional tail distributions 
$P(d>\ell | {\rm trap})$
and
$P(d>\ell | {\rm ret})$
of first hitting times 
vs. $\ell$,
for NBWs on an ER network
of size $N=1000$ and $c=3$, $5$ and $7$. 
The theoretical results,
obtained from Eqs.
(\ref{eq:CondTailTrap})
and
(\ref{eq:CondTailRet})
are found to be in excellent agreement
with the numerical simulations.

Finally, we use Bayes' theorem to obtain the conditional
distributions:

\begin{equation}
\fl
{ P\left({\rm trap}|d>\ell\right)=
\left\{
\begin{array}{ll}
\sum\limits_{t=\ell+1}^{S-1}
e^{-c\left(t\right)}
\frac{P\left(d>t-1\right)}{P\left(d>\ell\right)}
+
e^{-c(S-1)}
\sum\limits_{t=S}^{N}
\frac{P\left(d>t-1\right)}{P\left(d>\ell\right)}
& \ell \le S  
\\
e^{-c\left(S-1\right)}
\sum\limits_{t=\ell+1}^{N}
P\left(d>t-1\right)
& \ell\ge S+1
\end{array} 
\right.
}
\label{eq:PtrapS2}
\end{equation}

\noindent
and

\begin{equation}
\fl
{ P\left( {\rm ret}|d>\ell\right)=
\left\{
\begin{array}{ll}
\frac{\left[ 1-e^{-c\left(S-1\right)} \right]}{P\left(d>\ell\right)} 
\sum\limits_{t=S}^{N}
\left[\frac{c-c\left(t\right)-Sp}{c-(S+1)p+1}\right]
{ P\left(d>t-1\right) } 
& \ell \le S
\\
\frac{\left[ 1-e^{-c\left(S-1\right)} \right]}{P\left(d>\ell\right) }
\sum\limits_{t=\ell}^{N-2}
\left[\frac{c-c\left(t\right)-Sp}{c-(S+1)p+1}\right] 
{ P\left(d>t-1\right) }
& \ell\ge S+1
\end{array} 
\label{eq:PretS2}
\right.
}
\end{equation}

\begin{figure}
\centerline{
\includegraphics[width=18cm]{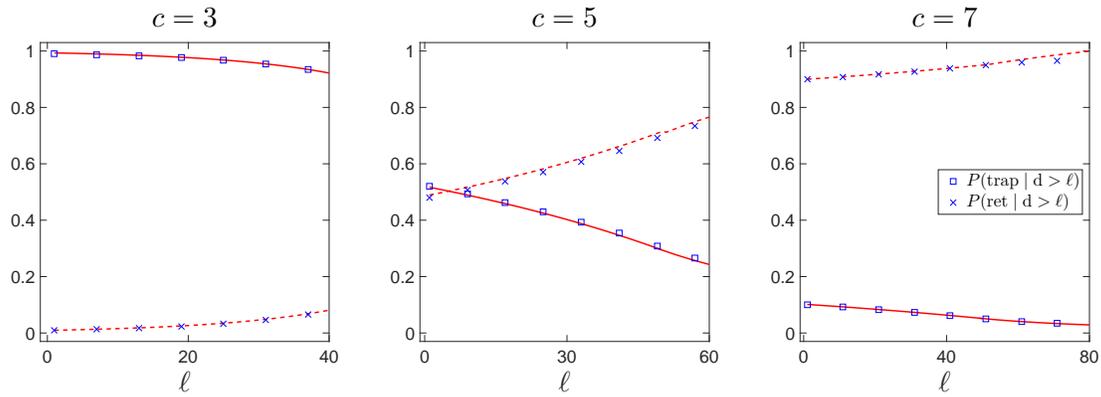}
}
\caption{
The conditional probabilities
$P({\rm trap} | d>\ell)$
and
$P({\rm ret} | d>\ell)$
that a path of a generalized NBW with $S=50$ will terminate by trapping or by retracing,
respectively, given that its length is larger than $\ell$,
are presented as a function of $\ell$.
Results are shown for ER networks of size $N=1000$ and 
$c=3$, $5$ and $7$.
The theoretical results for
$P({\rm trap} | d>\ell)$ (solid lines)
are obtained from Eq.
(\ref{eq:PtrapS2})
while the theoretical results for
$P({\rm ret} | d>\ell)$ (dashed lines)
are obtained from Eq.
(\ref{eq:PretS2}).
The theoretical results are compared to the results of numerical
simulations (symbols) finding excellent agreement. 
It is found that 
$P({\rm trap} | d>\ell)$
is a monotonically decreasing function of $\ell$
while
$P({\rm ret} | d>\ell)$ 
is monotonically increasing.
}
\label{fig:15}
\end{figure}

\noindent
In Fig.
\ref{fig:15}
we present the
conditional probabilities
$P({\rm trap} | d>\ell)$
and
$P({\rm ret} | d>\ell)$
that an NBW path will terminate by trapping or by retracing,
respectively, given that its length is larger than $\ell$,
as a function of $\ell$.
Results are shown for ER networks of size $N=1000$ and 
$c=3$, $5$ and $7$.
The theoretical results, 
obtained from Eqs.
(\ref{eq:PtrapS2})
and
(\ref{eq:PretS2})
are found to be in excellent agreement with the numerical simulations.
This comparison
is done for the range of path lengths which actually appear in the
numerical simulations and for which good statistics can be obtained.
Longer RW paths which extend beyond this range become extremely
rare, so it is difficult to obtain sufficient numerical data.
However, in the bottom row we show the theoretical results
for the entire range of path lengths. 
In fact, such long paths can be sampled using the pruned 
enriched Rosenbluth method, which was successfully used in the 
context of SAWs in polymer physics 
\cite{Grassberger1997}. 
In this method one samples long non-overlapping paths, 
keeping track of their weights, to obtain an unbiased sampling 
in the ensemble of all paths.

\begin{figure}
\centerline{
\includegraphics[width=8cm]{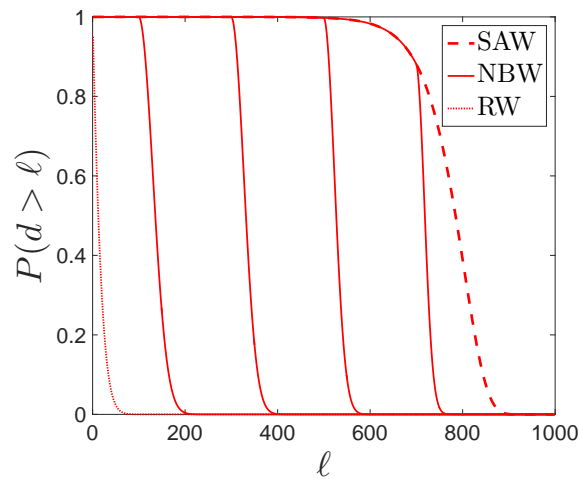}
}
\caption{
Theoretical results for the tail distributions
$P(d > \ell)$ 
of first hitting times of generalized NBWs
with $S=100$, $300$, $500$ and $700$
(solid lines, left to right),
in comparison with
the tail distribution of first hitting times 
of RWs (dotted line)
and the tail distribution
of last hitting times of SAWs
(dashed line),
on an ER network
of size $N=1000$ 
and 
$c=20$.
As $S$ is increased, the tail 
distribution of first hitting times
of the generalized NBWs exhibits a crossover from
the behavior of a RW towards the behavior of an SAW.
}
\label{fig:16}
\end{figure}

In Fig.
\ref{fig:16}
we present 
theoretical results for the tail distributions
$P(d > \ell)$ 
of first hitting times of generalized NBWs
with $S=100$, $300$, $500$ and $700$
(solid lines, left to right),
in comparison with
the tail distribution of first hitting times 
of RWs (dashed line)
and the tail distribution
of last hitting times of SAWs
(dotted line),
on an ER network
of size $N=1000$ 
and 
$c=20$.
As $S$ is increased, the form of the tail 
distribution exhibits a crossover from
the RW towards the SAW.

\section{Summary and Discussion}

NBWs provide useful insight on the dynamics of objects
which exhibit random motion on networks such as
web crawlers and foragers.
Starting from a random initial node,
these walkers hop randomly between adjacent nodes
without backtracking, namely without hopping back into
the previous node.
The NBW path terminates when it steps into 
a node which they already visited before (retracing scenario)
or when it becomes trapped in a dead-end node
from which it cannot exit (trapping scenario).
The number of steps taken from the initial node
up to the termination of the path is called the first
hitting time.
We obtained 
analytical results for the distribution 
of first hitting times,
$P(d=\ell)$,
of NBWs on ER networks
and for its mean, median and standard deviation.
We calculated the probabilities,
$p_{\rm trap}$ 
and 
$p_{\rm ret}$,
that an NBW path starting at a random node
will terminate by trapping or by retracing, respectively.
We also obtained analytical expressions for the conditional 
distributions of path lengths,
$P(d=\ell | {\rm trap})$ 
and
$P(d=\ell | {\rm ret})$
for the paths which terminate by
backtracking and by retracing,
respectively.
Finally, we calculated 
the conditional probabilities
$P({\rm trap} | d=\ell)$
and
$P({\rm ret} | d=\ell)$
that a path which terminates after $d=\ell$ steps
is terminated by trapping or by retracing, respectively.
It was found that the two termination mechanisms exhibit
very different behavior.
The trapping probability sets in starting from the 
second step and is constant throughout the path.
As a result, this mechanism alone would produce a 
geometric distribution of path lengths.
The retracing mechanisms sets in starting from the 
third step and its rate increases linearly in time.
The balance between the two termination mechanisms 
depends on the mean degree
of the network.
In the limit of sparse networks, 
the trapping mechanism is dominant and most
paths terminate long before the retracing mechanism becomes relevant.
In the case of dense networks, the trapping probability is low and most
paths terminate by the retracing mechanism.

Comparing the NBW model studied here 
to the RW model 
studied in Ref. 
\cite{Tishby2016b}, 
we find that
the probability of termination by retracing at time $t$,
given by
 
\begin{equation}
\fl
P_{\rm ret}(d=t | d>t-1) 
= 1-P_{\rm ret}(d>t | d>t-1)
=\frac{c-c(t)}{c+1},
\end{equation}

\noindent
is identical in the two models.
It can be expressed in the form

\begin{equation}
\fl
P_{\rm ret}(d=t | d>t-1) = 
\left( \frac{c}{c+1} \right) \left( \frac{t}{N-1} \right).
\end{equation}

\noindent
The difference between the distributions of first hitting times
in the two models is due to the different behaviors of the backtracking
mechanism in the RW model
and the trapping mechanism in the NBW model. 
While the probabilities of both backtracking
and trapping do not depend on time, they depend differently on the parameter $c$.
The backtracking probability is given by
\cite{Tishby2016b}

\begin{equation}
\fl 
P_{\rm backtrack}(d=t | d>t-1) =
1-P_{\rm backtrack}(d>t | d>t-1) = 
\frac{1-e^{-c}}{c},
\end{equation}

\noindent
while the trapping probability is given by

\begin{equation} 
\fl
P_{\rm trap}(d=t | d>t-1) =1-P_{\rm trap}(d>t | d>t-1) = e^{-c}.
\end{equation}

\noindent
This means that the backtracking 
probability essentially decreases as $1/c$
while the trapping probability decreases exponentially. 
Therefore, the trapping mechanism is 
much less likely to occur and the 
NBW paths are are much longer than the corresponding RW paths.

\begin{figure}
\centerline{
\includegraphics[width=8cm]{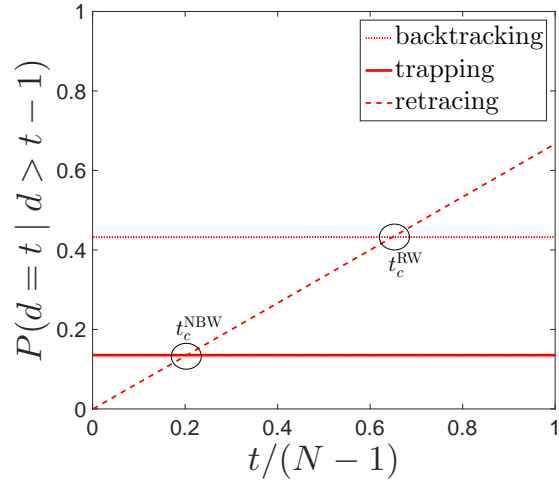}
}
\caption{
The probability
$P_{\rm backtrack}(d=t|d>t-1)$
that a RW will terminate by backtracking at time $t$
(dotted line),
the probability
$P_{\rm trap}(d=t|d>t-1)$
that an NBW will terminate by trapping at time $t$
(solid line)
and the probability
$P_{\rm ret}(d=t|d>t-1)$
that either a RW or an NBW will terminate by retracing at time $t$
(dashed line),
vs. the normalized time, $t/(N-1)$,
for an ER network of $N=1000$ nodes and $c=2$.
While the backtracking and trapping probabilities do not depend on
time, the retracing probability increases linearly with time.
Since the backtracking probability is larger than the trapping
probability, the crossover time
$t_c^{\rm RW}$
in which the retracing probability exceeds the backtracking
probability is larger than the crossover time
$t_c^{\rm NBW}$
in which the retracing probability exceeds the trapping probability.
}
\label{fig:17}
\end{figure}

In Fig.
\ref{fig:17}
we present the termination probabilities 
by backtracking (of RWs), by trapping (of NBWs) and by retracing
(both RWs and NBWs) as a function of time,
on an ER network of size $N=1000$ and $c=2$.
While the backtracking and trapping 
probabilities do not depend on time,
the retracing probability increases linearly with time.
The RW model exhibits a crossover time, 
$t_c^{\rm RW}$,
at which the retracing probability 
exceeds the backtracking probability.
This crossover time is given by

\begin{equation}
\frac{t_c^{\rm RW}}{N-1} = 
\left(\frac{c+1}{c}\right)
\left(\frac{1 - e^{-c}}{c} \right) 
\end{equation}

\noindent
Similarly, the NBW model exhibits
a crossover time,
$t_c^{\rm NBW}$,
given by

\begin{equation}
\frac{ t_c^{\rm NBW} }{N-1} = 
\left(\frac{c+1}{c}\right)e^{-c},
\end{equation}

\noindent
at which the retracing probability 
exceeds the trapping probability.

\begin{figure}
\centerline{
\includegraphics[width=8cm]{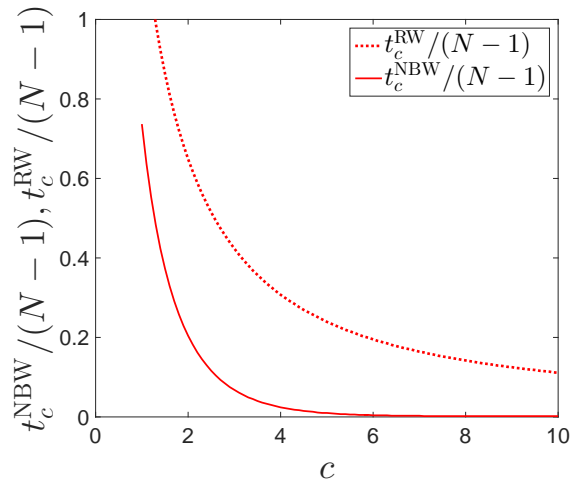}
}
\caption{
The normalized crossover time,
$t_c^{\rm RW}/(N-1)$,
in which the retracing probability exceeds the backtracking
probability of the RW
(dashed line)
and the normalized crossover time
$t_c^{\rm NBW}/(N-1)$
(solid line)
at which the retracing probability exceeds the trapping probability
of the NBW, as a function of $c$ for an ER network of size $N=1000$.
Both crossover times decrease as $c$ is increased,
while
$t_c^{\rm RW} > t_c^{\rm NBW}$
for all values of $c$.
}
\label{fig:18}
\end{figure}

In Fig.
\ref{fig:18}
we present the normalized
crossover times
$t_c^{\rm RW}/(N-1)$ 
and
$t_c^{\rm NBW}/(N-1)$
of RWs and NBWs, respectively, 
as a function of the mean degree, $c$,
on an ER network of size $N=1000$.
While both times are decreasing functions
of $c$, the crossover time of the RW
is larger than the crossover time of the NBW 
for all values of $c>1$.

We also considered generalized NBWs 
which keep track of the last $S$ visited
nodes and avoid hopping into them.
The parameter, $S$, may take values in the range
$1 \le S \le N-1$.
The case of $S=1$ is the NBW model,
in which backtracking into the previous node is not allowed.
The paths of NBWs may terminate
either by hopping into nodes already visited at earlier times (retracing)
or by trapping in a leaf node of degree $k=1$ from which they
cannot exit.
Generalized NBW models, with
$S>1$, avoid hopping into
any of the nodes visited at times $t=1$, $t-2$, $\dots$, $t-S$.
The paths of generalized NBWs may terminate either by retracing 
or by trapping in a node which
is surrounded by the tail of $S$ nodes which they cannot enter.
Thus, unlike the NBW which may be trapped only at a node of degree
$k=1$, the generalized NBW may be trapped at nodes of degrees
$k \le S$.
The case of $S=N-1$ coincides with the SAW model.
During the first $S+1$ time steps, the generalized NBW behaves like
an SAW and may terminate only by trapping. Starting at the $S+2$ step,
the retracing mechanism is activated and from that point on its weight
increases linearly with time.
Thus, the probability of termination by trapping $p_{\rm trap}$ is an
increasing function of $S$, while the probability of termination by
retracing $p_{\rm ret}$ is a decreasing function of $S$.
Overall, the mean path length of the generalized NBW increases
as $S$ is increased.

The analysis presented in this paper provides useful insight on
the broad issue of life-time distributions such as
the distribution of life expectancies of humans and animals 
and the distribution of service lives of machines
\cite{Finkelstein2008,Gavrilov2001}.
In particular, the RW, NBW and SAW provide well defined termination
mechanisms for which the distributions of termination times can be
calculated analytically. In case of the SAW, the termination rate increases
exponentially with time, giving rise to a distribution of termination times
which follows the Gompertz distribution.
In case of the RW and NBW, the termination rate consists of a constant
term and a linearly increasing term, resulting in a distribution of termination
times which follows a combination of 
an exponential distribution and a Rayleigh distribution. 
The termination rate of the generalized NBW resembles an SAW
in the first $S$ steps and an NBW afterwards. 
Interestingly, the life expectancies of humans can be described by the
Gompertz distribution 
\cite{Gompertz1825,Johnson1995,Shklovskii2005,Ohishi2008},
while the service lives of machines can be fitted to
a Weibull distribution, of which the Rayleigh distribution is a special case
\cite{Erumban2008}.
In general, the distribution of human life expectancies and the distribution
of machine service lives are determined by a combination of different causes
of death or failure mechanisms. To understand the interplay between these
different scenarios one needs to be able to isolate the contribution of each
of of them and examine how it varies with age. 
In the analysis of the termination scenarios of NBW paths we develop an
approach which enables us to disentangle the contribution of the trapping
and retracing mechanisms.
This type of analysis is likely to be useful in many other contexts in which
several failure mechanisms coexist.

\appendix

\section{Calculation of $P(d>\ell)$ for the NBW model}

To obtain a closed form expression for the tail distribution, 
$P(d>\ell)$,
we
take the natural 
logarithm on both sides of Eq.
(\ref{eq:cond2}).
This leads to

\begin{equation}
\ln \left[P\left(d>\ell\right)\right] =
\ln \left[P_{\rm ret}\left(d>\ell\right)\right]+
\ln \left[P_{\rm trap}\left(d>\ell\right)\right].
\label{eq:logtail1}
\end{equation}

\noindent
Below we analyze separately 
each one of the two terms on the right hand side
of Eq.
(\ref{eq:logtail1}).
The calculation of the tail distribution
$P_{\rm trap}(d>\ell)$ 
is simplified by the fact that 
$P_{\rm trap}(d>t | d>t-1)$
does not depend on $t$.
As a result, Eq.
(\ref{eq:cond4})
can be written in the form

\begin{equation}
P_{\rm trap}(d>\ell) 
= \left(1 - e^{-c} \right)^{\ell},
\label{eq:tailb}
\end{equation}

\noindent
or in the form

\begin{equation}
P_{\rm trap}(d>\ell) = e^{- \beta \ell},
\end{equation}

\noindent
where

\begin{equation}
\beta = - \ln \left( 1-e^{-c} \right).
\label{eq:betaA}
\end{equation}

\noindent
The termination by the trapping scenario can be considered as a 
Poisson process, in which
the termination probability  
is fixed and depends only on the mean
degree of the network.

Taking the logarithm of 
$P_{\rm ret}(d>\ell)$,
as expressed in Eq.
(\ref{eq:cond6}),
we obtain

\begin{equation}
\ln \left[P_{\rm ret}(d>\ell)\right] = 
\sum_{t=2}^{\ell} \ln \left[ \frac{c(t)-p+1}{c-2p+1} \right].
\label{eq:P_rsum}
\end{equation}

\noindent
Replacing the sum by an integral we obtain

\begin{equation}
\ln P_{\rm ret}(d>\ell)
\simeq
\int_{3/2}^{\ell+1/2} 
\ln \left[\frac{c\left(t\right)-p+1}{c-2p+1}\right]dt.
\label{eq:P_rint}
\end{equation}

\noindent
Plugging in the expression for $c(t)$ 
from Eq.
(\ref{eq:coft})
and rearranging terms in the integrand
we obtain

\begin{equation}
\ln P_{\rm ret}(d>\ell)
\simeq
\int_{3/2}^{\ell+1/2}
\ln \left[1-\frac{(t-1)c}{\left(N-1\right)
\left(c-2p+1\right)}\right]dt.
\label{eq:P_r3}
\end{equation}

\noindent
For sufficiently large networks 
one can replace
$N-1$ 
by 
$N$ 
and 
$c+1-2p$ 
by 
$c+1$.
Solving the integral we obtain

\begin{eqnarray}
\ln [P_{\rm ret}(d>\ell)]
&\simeq&
 \left(\ell-\frac{1}{2}-\alpha^2 \right) 
\ln \left(1 - \frac{\ell-1/2}{\alpha^2}\right)-(\ell-1)
\nonumber \\
&+&
\left(\alpha^2 - \frac{1}{2} \right) 
\ln \left(1 - \frac{1}{2\alpha^2}\right),
\label{eq:P_r4}
\end{eqnarray}

\noindent
where

\begin{equation}
\alpha=\sqrt{\frac{N\left(c+1\right)}{c}}.
\label{eq:alpha2}
\end{equation}

In the approximation of the sum of 
Eq. (\ref{eq:P_rsum}) by the integral of Eq. (\ref{eq:P_rint})
we have used the
formulation of the middle Riemann sum. Since the function
$\ln[P_r(d>\ell)]$ is a monotonically decreasing function, the value of
the integral is over-estimated by the left Riemann sum, $L_{\alpha}(\ell)$, and under-estimated
by the right Riemann sum, $R_{\alpha}(\ell)$. The error involved in this approximation is thus 
bounded by the difference $\Delta_{\alpha}(\ell) = L_{\alpha}(\ell) - R_{\alpha}(\ell)$,
which satisfies
$\Delta_{\alpha}(\ell) < \ln(1 - \ell/\alpha^2)$.
Thus, the relative error in $P(d>\ell)$ due to the approximation of the sum
by an integral is bounded by
$\eta_{\rm SI} = \ell/\alpha^2$,
which scales like $\ell/N$. 
Comparing the values obtained from the sum and the integral
over a broad range of parameters, we find that the pre-factor of
the error is very small, so in practice the error introduced by approximation of the
sum by an integral is negligible.

\section{Calculation of $P(d>\ell)$ for the generalized NBW model when $\ell > S$}

For $\ell > S$, 
the probability that the path length of the NBW will
be longer than $\ell$ is given by

\begin{equation} 
P(d>\ell) = P_{\rm ret}(d>\ell) P_{\rm trap}(d>\ell),
\label{eq:cond}
\end{equation}

\noindent
where

\begin{equation} 
P_{\rm trap}(d>\ell) = 
\prod_{t=1}^{\ell} P_{\rm trap}(d > t|d > t-1)
\label{eq:cond24}
\end{equation}

\noindent
and

\begin{equation} 
P_{\rm ret}(d>\ell) = 
\prod_{t=1}^{\ell} P_{\rm ret}(d > t|d > t-1).
\label{eq:cond25}
\end{equation}

The probability $P_{\rm trap}(d>\ell)$ 
is given by 

\begin{equation}
P_{\rm trap} \left(d>\ell\right) =
\left(
\prod_{t=1}^{S-1}
\left[1-e^{-c\left(t\right)}\right]
\right)
\left[1-e^{-c\left(S-1\right)}\right]^{\ell_S},
\label{eq:P_t_t>S}
\end{equation}

\noindent
where

\begin{equation}
\ell_{S}=\ell-(S-1),
\label{eq:ell_S}
\end{equation}

\noindent
and 
$c(S-1)$ 
and
$c(t)$ 
are given by Eq.
(\ref{eq:coft}).

The probability $P_{\rm ret}(d>\ell)$ can be written in the form

\begin{equation} 
P_{\rm ret}(d>\ell) =  \prod_{t=S}^{\ell} 
\left[ \frac{c(t)-p+1}{c-(S+1)p+1} \right].
\label{eq:cond6S}
\end{equation}

\noindent
Taking the logarithm of $P_{\rm ret}(d>\ell)$
we obtain

\begin{equation}
\ln \left[P_{\rm ret}(d>\ell)\right] = 
\sum_{t=S+1}^{\ell} \ln \left[ \frac{c(t)-p+1}{c-(S+1)p+1} \right].
\end{equation}

\noindent
Replacing the sum by an integral, 
plugging in the expression for $c(t)$ 
from Eq.
(\ref{eq:coft})
and rearranging terms in the integrand,
we obtain

\begin{equation}
\ln P_{\rm ret} \left(d>\ell\right) = 
\int_{S+1/2}^{\ell+1/2}\ln\left[1-\frac{\left(t-S\right)c}{\left(N-1\right)\left(c-2p+1\right)}\right]dt.
\end{equation}

\noindent
Changing the integration variable to
$t^{\prime} = t - S + 1$,
the lower limit of the integration becomes
$t^{\prime} = 3/2$,
while the upper limit becomes
$t^{\prime} = \ell_S + 1/2$.
The integral takes the form

\begin{equation}
\ln P_{\rm ret}\left(d>\ell\right) = 
\int_{3/2}^{\ell+1/2}
\ln\left[ 1 - \frac{(t'-1)c}{\left(N-1\right)\left(c-2p+1\right)} \right] dt'.
\end{equation}

\noindent
Solving this integral we obtain

\begin{eqnarray}
\ln [P_{\rm ret}(d>\ell)]
&\simeq&
 \left(\ell_S-\frac{1}{2}-\alpha^2 \right) 
\ln \left(1 - \frac{\ell_S-1/2}{\alpha^2}\right)-(\ell_S-1)
\nonumber \\
&+&
\left(\alpha^2 - \frac{1}{2} \right) 
\ln \left(1 - \frac{1}{2\alpha^2}\right)
+1,
\label{eq:P_r4A}
\end{eqnarray}

\section{Calculation of $\ell_{\rm mean}$ for the generalized NBW model}

Since 
$P\left(d>0\right)=1$, 
the mean path length can be expressed in the form

\begin{equation}
\ell_{\rm mean} =
1 + I_1 + I_2,
\label{eq:ell_mmeanS1}
\end{equation}

\noindent
where

\begin{equation}
I_1 = 
\sum_{\ell=1}^{S}P_{\rm trap}\left(d>\ell\right),
\end{equation}

\noindent
and

\begin{equation}
I_2 = 
\sum_{\ell=S+1}^{N-2}P\left(d>\ell\right).
\end{equation}

\noindent
Replacing the sums by integrals we obtain

\begin{equation}
I_1 \simeq 
\int_{\frac{1}{2}}^{S+\frac{1}{2}}P_{\rm trap}\left(d>\ell\right) d\ell,
\label{eq:ell_meanS11}
\end{equation}

\noindent
and

\begin{equation}
I_2 \simeq 
\int_{S+\frac{1}{2}}^{N-\frac{3}{2}}P\left(d>\ell\right) d\ell.
\label{eq:ell_meanS12}
\end{equation}

\noindent
Inserting $P_{\rm trap}(d>\ell)$ from Eq.
(\ref{eq:tail2})
into Eq.
(\ref{eq:ell_meanS11})
we obtain the integral

\begin{equation}
I_1 \simeq
\int_{\frac{1}{2}}^{S+\frac{1}{2}}
\exp\left[-\eta \left(e^{b\ell}-1\right)\right]d\ell.
\end{equation}

\noindent
Its solution is given by

\begin{equation}
I_1 \simeq
\frac{e^{\eta}}{b}
\left[{\rm Ei}\left( -\eta e^{b(S+{1}/{2})} \right)
-{\rm Ei}\left(-\eta e^{ \frac{b}{2}} \right)\right].
\end{equation}

\noindent
The integral $I_2$ can be written in the form

\begin{equation}
I_2
\simeq
\exp\left[-\eta \left(e^{b S}-1\right)\right]
\int_{\frac{1}{2}}^{N-S-\frac{3}{2}}
\exp\left(-\frac{\ell_{S}(\ell_S - 1)}{2\alpha^2}
-\beta_{S}\ell_{S}\right)d\ell_{S}.
\end{equation}

\noindent
Following Ref.
\cite{Tishby2016b}
we obtain

\begin{eqnarray}
I_2 &=& 
\sqrt{\frac{\pi}{2}} \alpha
\exp\left[\frac{(\alpha^2 \beta_{S} - 1)\beta_S}{2} 
- \eta \left(e^{b S}-1\right)\right]
\nonumber \\
&\times&
\left[{\rm erf}\left(\frac{\alpha^2\beta_{S}+(N-S)-2}{\sqrt{2}\alpha}\right)
-{\rm erf}\left(\frac{\alpha\beta_{S}}{\sqrt{2}} \right)\right].
\end{eqnarray}

\noappendix

\end{document}